\documentclass[a4paper,12pt,twoside,openright,notitlepage]{report}
\input{english.sty}
\usepackage{a4}
\usepackage{fancyheadings}
\usepackage{amssymb}
\usepackage{amsbsy}
\usepackage{amsmath}

\newcommand{\be}{\begin{equation}} \newcommand{\ee}{\end{equation}}
\newcommand{\bea}{\begin{eqnarray}} \newcommand{\eea}{\end{eqnarray}}
\newcommand{\el}{\nonumber \\}
\newcommand{\re}[1]{(\ref{#1})}
\newcommand{\pat}{\partial}
\newcommand{\adot}{\dot{a}} 
\newcommand{\bdot}{\dot{b}} \newcommand{\bddot}{\ddot{b}}

\newcommand{\Ldot}{\dot{L}}

\newcommand{\rhodot}{\dot{\rho}}

\newcommand{\Ydot}{\dot{Y}} 
\newcommand{\at}{\tilde{a}} \newcommand{\nt}{\tilde{n}}
\newcommand{\atdot}{\dot{\at}}
\newcommand{\phit}{\tilde{\phi}}
\newcommand{\phitdot}{\dot{\phit}}

\begin{document}
\baselineskip16pt
\setcounter{footnote}{0}

\begin{titlepage}
\begin{flushleft}
       \hfill                      {\tt astro-ph/0208282}\\ \hfill
       HIP-2002-05 \\ \hfill            August 14, 2002\\
\end{flushleft}
\vspace*{3mm}
\begin{center}
{\Large {\bf A primer on the ekpyrotic scenario}\\}
\vspace*{12mm} {\large Syksy R\"{a}s\"{a}nen\footnote{Email: syksy.rasanen@iki.fi}\\}

\vspace{5mm}

{\em {}Helsinki Institute of Physics \\ P.O. Box 64,
FIN-00014 University of Helsinki, Finland\footnote{After October 1, 2002: {\em {} Theoretical Physics, University of Oxford, 1 Keble Road,
Oxford OX1 3NP, UK}}}

\vspace{3mm}

\vspace*{10mm}

\end{center}

\begin{abstract}

\noindent This is an introduction to the ekpyrotic scenario, with
an emphasis on the two contexts of brane cosmology and primordial
universe scenarios. A self-contained introduction to
brane cosmology and a qualitative overview and comparison of the
inflationary, pre-big bang and ekpyrotic scenarios are given as
background. The ekpyrotic scenario is then presented in more detail,
stressing various problems.

\end{abstract}

\end{titlepage}

\pagestyle{fancy}

\tableofcontents

\setcounter{footnote}{0}

\setcounter{secnumdepth}{3}

\chapter{Introduction} \label{intro}

\section{An overview of the paper}

This is intended to be a useful general level introduction
to the \emph{ekpyrotic scenario}
[1-7].
Familiarity with brane cosmology topics such as the
\emph{Randall-Sundrum model} \cite{Randall:1999a, Randall:1999b}
or other related fields is not assumed. I have attempted to keep the
presentation reasonably self-contained, and it should be up-to-date
on references as regards the ekpyrotic scenario (though not necessarily
on tangential topics).

The ekpyrotic scenario was presented about a year and a half ago, in March
2001, as an alternative to the prominent scenarios of the primordial universe,
inflation and pre-big bang. The aim of the scenario is to provide
solutions to major cosmological problems, including the homogeneity and
isotropy problem, the flatness problem and the problem of the seeds of
large-scale structure, on the basis of fundamental physics.

The starting point of the ekpyrotic scenario is the unified theory
known as \emph{heterotic M-theory}
[10-12]. This theory is eleven-dimensional, and the compact eleventh
dimension is bounded at both ends by ten-dimensional slices known as
\emph{branes}\footnote{These are not the same objects as the D-branes of
string theory.}. These branes play a vital role in heterotic M-theory.
However, in the context of cosmology such codimension one objects have
mostly been investigated in phenomenological constructions, such as the
Randall-Sundrum model, instead of the fundamental heterotic M-theory. The
understanding gained by studying these phenomenological constructions can to
a large degree be applied to the ekpyrotic scenario, so that
phenomenological brane cosmology provides an important background for the
ekpyrotic scenario in addition to heterotic M-theory.

Chapter \ref{rs} puts brane physics into the historical context
of extra dimension theories and contains an overview of the basics
of the brane scenario. It starts from the Randall-Sundrum model
and proceeds to the general case of brane gravity and cosmology
in the case of one extra dimension. The main result of studies of
brane gravity is emphasised: it is possible to obtain
approximately four-dimensional gravity independent of the size of the
extra dimension, in contrast to set-ups where the observers are not
localised in the extra dimension.

Chapter \ref{cosmo} discusses the cosmological background.
The main present cosmological problems are listed, and the solutions
offered by the most studied comprehensive scenarios of the
primordial universe --inflation and pre-big bang-- as well as
by the ekpyrotic scenario are presented, along with the problems of
the well-studied scenarios.

Chapter \ref{ekpyrotic} contains a more detailed presentation of the
ekpyrotic scenario, including its problems. The heterotic M-theory set-up
is presented and the construction and analysis of the four-dimensional
effective theory is outlined.  After briefly discussing the internal
problems of the four-dimensional effective theory, the far more serious
problems of the four-dimensional construction itself are addressed.
Some problems faced by the five-dimensional approach are then discussed,
and their relevance to the so-called ``cyclic model of the
universe'' 
[13-15] --a spin-off of the ekpyrotic scenario-- is commented upon.

I conclude that the ekpyrotic scenario is a welcome new idea but that
most work done thus far is not solid. Careful analysis in the
higher-dimensional setting is needed to promote the scenario from an
interesting concept to a working model with testable predictions.

This paper is a revised version of the introductory part of my Ph.D.
thesis at the Department of Physical Sciences at the University of Helsinki;
the thesis consists of the introductory part and the publications
\cite{Enqvist:2000, Enqvist:2001a, Rasanen:2001}
The original version of the introductory part, titled \emph{Topics in brane
cosmology,} is available at http://ethesis.helsinki.fi.
I have added some references and comments, fixed a few typos in the text and
made some other, mostly trivial, changes. The only significant change is that a
mistake regarding the constraints on brane matter in boundary
brane--boundary brane collisions in section \ref{collision} (and in
\cite{Rasanen:2001}) has been corrected.

\section{Acknowledgments}

I wish to thank my thesis supervisors Kari Enqvist and Esko
Keski-Vakkuri. Kari, for lending his physics intuition and for tutoring
in the pragmatical aspects of reseach and publishing, as well as
for encouraging (and arranging funding for) me to attend numerous
conferences. Esko, for more involved guidance and many discussions.

Thanks are also extended to Fawad Hassan for explanations of
various aspects of string theory and useful discussions, Hannu
Kurki-Suonio for help with CMB and LSS references, Martin Sloth
for discussions and explanations on the pre-big bang scenario as well
as comments on the manuscript, Riccardo Sturani for answering my
questions on the pre-big bang scenario and Jussi V\"{a}liviita for
similarly clarifying things regarding CMB perturbations. Pikkukettu
Hopeakettu deserves special mention for many discussions and comments
on the manuscript.

It is a pleasure to acknowledge Robert Brandenberger's
pleasant yet firm conduct as an opponent at the thesis defense, as well as
his encouraging statements.

I wish to thank Andrei Linde, Burt Ovrut, Paul Steinhardt
and Neil Turok for discussions related to the ekpyrotic scenario.
The discussions played an important part in selecting both the subject
of my second paper on the ekpyrotic scenario and the contents of the
ekpyrotic parts of the thesis. Also, I wish to thank the referees of the
thesis, Jorma Louko and Claus Montonen, for their useful comments and
corrections.

Of course, all errors, omissions and opinions are mine and not to be
blamed on others.

Conferences have been important to my research, and the trips have been
supported by the Magnus Ehrnrooth Foundation and the Jenny and Antti
Wihuri Fund, to which I express my gratitude. The stipend of the
Waldemar von Frenckell foundation is also gratefully acknowledged.

I want to thank Christofer Cronstr\"{o}m for supervising the latter stages
of my Master's thesis in 1998-1999. Finally, I extend especially warm thanks
to Jouni Niskanen for encouraging me and for offering me a position at the
Theory Division of the Department of Physics after the behaviour of a
professor at the Department had left me in a state of shock. It is quite
possible that without Jouni's support I would not have continued doing
physics after finishing my Master's thesis.

I am beyond gratitude to Heidi Hopeamets\"{a} for her continued love and
support.

\chapter{The brane scenario} \label{rs}

\section{A brief overview of extra dimensions}

\subsection{A micro-history}

Spacetime dimensions additional to the observed four were introduced
to physics by Gunnar Nordstr\"{o}m in 1914 \cite{Nordstrom:1914}, and became
popular from the work of Kaluza and Klein in the 1920s
[20-22].
The early work on Nordstr\"{o}m-Kaluza-Klein
theories stemmed from a desire to unify gravity and electromagnetism in a
higher-dimensional, purely geometric framework. Since the theories
attempted to unify the most fundamental interactions of the day, they were
naturally phenomenological in character. For example, there was no principle
to dictate the number of extra dimensions.

A more well-founded framework for Kaluza-Klein theories was provided by
the advent of supergravity in 1976
[23-25].
First of all, supergravity sets a maximum limit of 11 on the number of
spacetime dimensions. This limit arises in the following way.  It is
believed (though not proven) that there are no consistent interacting
field theories with particles of spin more than two
[26-28].
The maximum helicity range is therefore from -2 to 2, implying a
maximum of 8 supersymmetry generators, since each changes helicity by
one-half. This requirement on the amount of symmetry in turn sets the
upper limit of 11 on the number of spacetime dimensions\footnote{Assuming
there is only one temporal direction.}. Second, besides
guiding the choice of the number of spacetime dimensions, supergravity
restricts the field content. The more symmetric a theory is, the more
constrained it is; in the maximum eleven dimensions supergravity is
unique, greatly increasing its appeal \cite{Duff:1986}.

While supergravity theories provide a well-motivated and potentially
realistic framework in which to implement Kaluza-Klein ideas, the extra
dimensions are still something of a luxury rather than a necessity:
aesthetic considerations may suggest more than four dimensions, but there is
nothing in supergravity theories that would require them. In this sense,
string theories opened a new era in (among other things) the field of extra
dimensions in 1974 \cite{Scherk:1974}. Bosonic string theory is consistent
only in 26 dimensions, and the introduction of supersymmetry brings the
number down to 10 for superstring theory\footnote{It is not impossible to
formulate string theory in less than 26 or 10 dimensions, but the
construction is less straightforward. For examples, see the covariant
lattice approach in \cite{Lust:1989} and the Landau-Ginzburg models and the
Gepner models in \cite{Polchinski:1998}.}. Different string theories are
believed to be different limits of a fundamental theory known as M-theory
that is formulated in one higher dimension, bringing the number to 11,
perhaps not incidentally the maximum for supergravity theories.

There are different formulations and limits of M-theory. The ekpyrotic
scenario is based on one known as heterotic M-theory
[10-12].
Heterotic M-theory describes the strong
coupling limit of heterotic string theory (which is a mixture of bosonic
string theory and superstring theory). The theory is formulated on an
orbifold ${\cal M}_{10}\times S_1/\mathbb{Z}_2$, where ${\cal M}_{10}$ is a
smooth ten-dimensional manifold. There are two points on the circle $S_1$
that are left under invariant under the action of the group $\mathbb{Z}_2$.
The smooth ten-dimensional manifolds at these \emph{fixed points} are called
branes and the space between the boundary branes is called the \emph{bulk}.

The low-energy limit of heterotic M-theory is eleven-dimensional
supergravity coupled to two ten-dimensional $E_8$ gauge theories, one on
each brane. Compactifying six spatial dimensions on a Calabi-Yau
manifold\footnote{Sometimes called a threefold in this context since it can
be expressed in terms of three complex dimensions.} leads to an effectively
five-dimensional gauged N=1 supergravity theory \cite{Lukas:1998a,
Lukas:1998b}. This five-dimensional theory with four-dimensional
contributions at the boundary of the orbifold ${\cal M}_{4}\times
S_1/\mathbb{Z}_2$ serves as the framework of the ekpyrotic scenario.
It will be presented in more detail in chapter \ref{ekpyrotic}.

\subsection{Where are they?}

An obvious problem in a physical theory with extra spacetime dimensions
is to reconcile their existence with observations, which to date are
all consistent with four spacetime dimensions. There are two kinds of
observations: those that measure the spacetime geometry directly via gravity
and those that measure it indirectly via observations of gauge
interactions. The approach introduced by Klein \cite{Klein:1926b},
followed in the early Kaluza-Klein theories as well as in supergravity
and string theories, is to take the extra dimensions to be compact and
small. The idea is that probes with wavelength much bigger than the size
of the extra dimension will not be able to resolve it, so that spacetime
looks four-dimensional at the low energies presently accessible to observation.

Gravity has been directly tested to distances of about .1 mm
\cite{Adelberger:2002}, while gauge interactions have been probed to
distances of the order of \mbox{(100 GeV)$^{-1}\sim10^{-18}$ m}
\cite{Groom:2001}. No deviations from the four-dimensional predictions have
been found. So, it seems that the size of extra dimensions in which fields
with gauge interactions propagate must be smaller than $10^{-18}$
m.\footnote{In the context of string theory this requirement is
easily fulfilled, since the natural scale of extra dimensions
is set by the string scale, which is usually within a few orders of
magnitude of the Planck scale
$M_{Pl}^{-1}\sim (10^{18} \textrm{GeV})^{-1}\sim 10^{-34}$ m. There are
also string models where the size of some extra dimensions is of the 
order of $(10^3 \textrm{GeV})^{-1}\sim 10^{-19}$ m \cite{Antoniadis:1990},
placing them on the threshold of experimental detection.} On the
other hand, this bound says nothing about dimensions in which fields with gauge
interactions do not propagate. An example would be the eleventh dimension in
heterotic M-theory, since gauge fields are confined to the ten-dimensional
boundary branes. However, since gravity is an expression of spacetime
geometry and thus by definition propagates in all dimensions, it might seem
that gravity experiments set a definite limit of the order of .1 mm on the
size of any extra dimensions. The main new contribution of the brane
scenario to the field of extra dimensions is to demonstrate that this is not
the case. When the observers are confined on a brane it is possible to
obtain the correct lower-dimensional gravity (within the current
observational limits) for large or even infinite extra dimensions.

\section{The Randall-Sundrum model}

\subsection{The basic construction}

Several papers discussing gravity in brane models appeared
in 1999
[37-40].
However, two publications by Randall and Sundrum
\cite{Randall:1999a, Randall:1999b} set off a voluminous exploration of
brane physics. Sometimes the term ``Randall-Sundrum model'' (or
``Randall-Sundrum scenario'') is used to refer to brane set-ups generally,
but I will take it to mean strictly the first model outlined by Randall
and Sundrum in their initial papers.

The framework of the proposal of Randall and Sundrum is general relativity
plus classical field theory. No quantum effects or unified theories are
involved, and in particular there is no reference to heterotic M-theory,
though the set-up is quite similar. The spacetime is taken to be a
five-dimensional orbifold \mbox{${\cal M}_{4}\times S_1/\mathbb{Z}_2$.} One of
the boundary branes is identified with the visible universe, also called
the \emph{visible brane,} and the other with a ``hidden universe'', or
\emph{hidden brane.} The branes are assumed to be parallel. Fields that feel
gauge interactions are assumed to be confined to the branes, so that
only gravity propagates in the extra dimension. There are cosmological
constants on the branes and in the bulk. The action is
\bea \label{rsaction}
  S_{\textrm{RS}} &=& \int_{{\cal M}_5} d^4 x dy \sqrt{-g}\left(\frac{M_5^3}{2} R-\Lambda\right) \el
  & & + \sum_{i=1}^2 \int_{{\cal M}_4^{(i)}}\!\! d^4 x\sqrt{-h^{(i)}} \left( \Lambda_i + {\cal L}_{\textrm{matter}(i)} \right) \ ,
\eea

\noindent where $M_5$ is the Planck mass in five dimensions, $R$ is the
scalar curvature in five dimensions, $\Lambda$, $\Lambda_1$ and $\Lambda_2$
are the cosmological constants in the bulk, on the visible brane and
on the hidden brane, respectively and ${\cal L}_{\textrm{matter}(i)}$ is
the Lagrange density of matter on brane $i$, including the Standard Model
fields on the visible brane. The coordinates $x^{\mu}$ and $y$ are the
coordinates parallel and perpendicular to the branes, respectively,
and $y$ covers the range from $0$ to $\pi r_c$, where $r_c$ is a
constant. The visible brane is at $y_1=\pi r_c$ and the hidden brane is
at $y_2=0$. The tensor $g_{AB}$ is the metric on ${\cal M}_5$ and
$h^{(i)}_{\mu\nu}=g_{\mu\nu}(x^{\mu}, y_i)$ are the induced metrics on
the branes ${\cal M}_4^{(i)}$.

The ``vacuum state'' of the above action, with four-dimensional
Poincar\'e invariance with respect to the dimensions parallel to the
brane (and ${\cal L}_{\textrm{matter}(i)}=0$) is a slice of anti-de
Sitter space,
\bea \label{rsmetric}
  ds^2 = e^{-2 k y}\eta_{\mu\nu}dx^{\mu} dx^{\nu} + d y^2 \ ,
\eea

\noindent where $\eta_{\mu\nu}$ is the four-dimensional Minkowski metric and
$k$ is a constant. The function $e^{-k y}$, and more generally (the
square root of) a function of $y$ multiplying the metric of the
four-dimensional spacetime spanned by $x^{\mu}$, is called the
\emph{warp factor}. The cosmological constants in the bulk and on the
branes are related to each other and to $k$ as follows:
\bea \label{lambda}
  \Lambda_1 &=& - \Lambda_2 \,=\, - 6 M_5^3 k \el
  \Lambda &=& - 6 M_5^3 k^2 \ ,
\eea

\noindent where $k$ is taken to be a few magnitudes below the Planck scale.
The spacetime curvature is thus near the Planck scale, so that it is not
clear how reliable the classical treatment of gravity is. Note that the
cosmological constant in the bulk and the cosmological constant on the
visible brane are negative, and that there is a fine-tuning between all
three cosmological constants.

It should be emphasised that the ``cosmological constants on the branes'' are
not the cosmological constants measured on the branes. In fact, the
fine-tuning \re{lambda} is equivalent to setting the observed
cosmological constants to zero, as is clear from the fact that
the geometry on the branes is flat and not (anti-)de Sitter. This is in part
because the relation between matter and curvature on the branes is
not given by the usual four-dimensional Einstein equation, and in part
because the bulk cosmological constant also contributes to brane gravity.
These issues will be considered in detail in section
\ref{branegravity}. The appropriate interpretation of the quantities
$\Lambda_i$ is that they are tensions related to the embedding of the branes
into the five-dimensional spacetime, so I will refer to them as ``brane
tensions'' from now on, reserving the term ``brane cosmological constant''
for the effective cosmological constant actually measured on the
brane\footnote{A useful convention also because in the ekpyrotic
scenario, the main object of interest, the brane tensions are not necessarily
constant but can vary in time; see \mbox{chapter \ref{ekpyrotic}.}}. 

\subsection{The hierarchy problem}

Though brane scenarios have interesting cosmological applications, the
motivation for the proposal of Randall and Sundrum did not stem from
cosmology but from particle physics. The model was intended to address the
so-called hierarchy problem, i.e. the origin of the vast difference between
the scale of gravity $M_{Pl}\sim 10^{18}$ GeV and the scale of electroweak
physics $\sim$ 100 GeV $\sim$ TeV. The Randall-Sundrum model addresses
the hierarchy problem in the following manner. Let us consider the action for
the Higgs field $H$
\bea \label{higgs}
  S_{\textrm{matter}(1)} &=& \int d^4 x {\cal L}_{\textrm{matter}(1)} \el
  & & \supset \int d^4 x\sqrt{-h^{(1)}}\left( h^{(1)\mu\nu}(D_{\mu}H)^{\dagger} D_{\nu}H - \lambda (\vert H\vert^2 - m_0^2)^2  \right) \ .
\eea

Substituting the ``vacuum'' metric \re{rsmetric} into \re{higgs}, we have
\bea
  S_{\textrm{matter}(1)} \supset \int d^4 x e^{-4 k\pi r_c}\left( e^{2 k\pi r_c}\eta^{\mu\nu}(D_{\mu}H)^{\dagger} D_{\nu}H - \lambda (\vert H\vert^2 - m_0^2)^2  \right) \ .
\eea

Making the field redefinition $H\rightarrow e^{k\pi r_c}H$, we obtain
\bea
  S_{\textrm{matter}(1)} \supset \int d^4 x \left(\eta^{\mu\nu}(D_{\mu}H)^{\dagger} D_{\nu}H - \lambda (\vert H\vert^2 - e^{-2 k\pi r_c} m_0^2)^2  \right) \ .
\eea

The ``bare'' mass $m_0$ has been replaced by a mass suppressed by the
warp factor, $m=e^{-k\pi r_c}m_0$. The same applies also to fields in
other representations of the Poincar\'e group, such as fermions.
Using the exponential warp factor one can generate an electroweak
scale mass $m\sim100$ GeV from a Planck scale mass $m_0\sim 10^{18}$ GeV
without the introduction
of large parameters; only $\pi k r_c\approx37$ is needed. From this
point of view, the smallness of the electroweak scale does not have a
particle physics origin but is an expression of the high curvature of
spacetime, which is in turn related to the Planck scale bulk cosmological
constant.

This resolution of the hierarchy problem relies on a small warp factor at the
visible brane. However, the normalisation of the warp factor has no physical
significance, and as pointed out by Randall and Sundrum, in the natural
coordinate system of an observer on the visible brane the warp factor
they measure is unity, $e^{-k (y-\pi r_c)}$. In this
coordinate system there is no rescaling of the ``bare'' masses in the visible
brane. However, one can reverse the view of hierarchy problem and say that
it is not the Planck scale but the electroweak scale which is fundamental.
Randall and Sundrum proposed that from this point
of view the particle masses are naturally of the electroweak scale, whereas
the large Planck mass arises from the exponential warp factor.

It is somewhat unclear how the emergence of Planck scale gravity from
the fundamental constants of the TeV scale is supposed to come about,
since the gravitational constant does not receive exponential
suppression. The four-dimensional Planck mass given by Randall and Sundrum is
\bea
  M_{Pl}^2 &=& \frac{M_5^3}{k} (1-e^{-2\pi k r_c}) \el
  &\approx& \frac{M_5^3}{k}
\eea

If the fundamental scale of gravity $M_5$ is of the TeV range, one needs
a large hierarchy, $k\approx M_5^3/M_{Pl}^2\sim 10^{-30}$ TeV, to get
the observed Planck mass. However, as a matter of fact the above
``Planck mass'' is not the Planck mass measured on the visible brane,
but instead a quantity which has been integrated over the fifth dimension.
The physical interpretation of such averaged quantities in a set-up
with observers strictly localised on a brane is not clear\footnote{This
issue will be discussed in more detail in section \ref{integ}.}.
Unfortunately, the physical Planck mass fares no better. The gravitational
coupling on the visible brane is \cite{Shiromizu:1999}
\bea \label{Mpl}
  8\pi G_N \equiv \frac{1}{M_{Pl}^2} = \frac{\Lambda_1}{6 M_5^6} \ ,
\eea

\noindent so that one needs $\Lambda_1\sim M_5^6 M_{Pl}^{-2}\sim 10^{-30}$
(TeV)$^4$, or $^4\sqrt{\Lambda_1}\sim 10^{-8}$ TeV, still a large
hierarchy. However, a more serious problem is visible in \re{Mpl}:
according to \re{lambda} the tension of the visible brane, $\Lambda_1$,
is negative, leading to a negative gravitational coupling. By switching
the assignment of the visible and hidden branes one obtains a positive
gravitational coupling, but then the warp factor is exponentially enhanced
instead of suppressed at the visible brane.

\section{Gravity on a brane} \label{branegravity}

\subsection{The general case}

Even if the Randall-Sundrum model does not offer solutions to
problems of particle physics, it may still lead to interesting 
modifications of gravity. There is a large literature on brane
gravity, particularly brane cosmology:
[1, 2, 4-9, 13-18, 37-67] offer some examples. Numerous publications and
preprints could be added, some more relevant than others.

Kaluza-Klein recipes for gravity are sometimes applied to brane cosmology.
However, apart from the questionable applicability of such
treatment\footnote{See section \ref{integ} for discussion.}, the
four-dimensional gravity in a brane setting is known exactly, so
that there is little need for such approximate methods. The exact
equations describing the four-dimensional gravity seen by an observer
on an infinitely thin brane in a generic five-dimensional setting were
derived in \cite{Shiromizu:1999}, assuming only the validity of general
relativity. The starting point is the Einstein equation in five
dimensions, written as
\bea \label{5deinstein}
  G_{AB} &=& \frac{1}{M_5^3} T_{AB} \ ,
\eea

\noindent where the Latin capital letters cover all five directions,
$G_{AB}$ is the five-dimensional Einstein tensor and $T_{AB}$ is the
five-dimensional energy-momentum tensor; the five-dimensional metric
is denoted by $g_{AB}$ as before. Note that $G_{AB}$ and $T_{AB}$
include both bulk and brane quantities, the latter as delta function
contributions. Given \re{5deinstein}, the induced Einstein equation on a
brane at a fixed point of $\mathbb{Z}_2$-symmetry  is \cite{Shiromizu:1999}
\bea \label{inducedgen}
  ^{(4)} G_{\mu\nu} &=& \frac{2}{3 M_5^3}\left( T_{AB}{}^{(4)}g^{A}_{\ \mu}\,^{(4)}g^{B}_{\ \nu} + \left( T_{AB}n^A n^B-\frac{1}{4}T^{A}_{\ A} \right) {}^{(4)}g_{\mu\nu}\right) \el
  && + \frac{1}{24 M_5^6} \bigg( -6 ^{(4)}T_{\mu\alpha}\ ^{(4)}T^{\alpha}_{\ \nu} + 2 ^{(4)}T^{\alpha}_{\ \alpha}{} ^{(4)}T_{\mu\nu} \el
  && + \left( 3 ^{(4)}T_{\alpha\beta}\ ^{(4)}T^{\alpha\beta} -{} ^{(4)}T^{\alpha}_{\ \alpha}{} ^{(4)}T^{\beta}_{\ \beta} \right) {}^{(4)} g_{\mu\nu} \bigg) - E_{\mu\nu} \ ,
\eea

\noindent where the Greek letters cover the directions parallel to the brane,
$n^A$ is a unit vector normal to the brane, $^{(4)}g_{\mu\nu}$ is the metric
induced on the brane (with $^{(4)}g_{AB}\equiv g_{AB}-n_A n_B$),
$^{(4)} G_{\mu\nu}$ is the Einstein tensor formed from the metric
$^{(4)}g_{\mu\nu}$, $^{(4)}T_{\mu\nu}$ is the energy-momentum tensor of the
brane and $E_{\mu\nu}$ is a traceless contribution related to the bulk Weyl
tensor $C_{ABCD}$ via
$E_{\mu\nu}=C_{ABCD} n^A n^C {}^{(4)}g^{B}_{\ \mu}\,^{(4)}g^{D}_{\ \nu}$.
Note that in the derivation of this result there are no constraints on
the energy-momentum tensors of the bulk or the visible brane, nor is there
any limitation on the number and matter content of possible other branes.

The induced Einstein equation \re{inducedgen} differs from the standard
four-dimensional Einstein equation in three respects. The dependence
of the four-dimensional Einstein tensor on the four-dimensional
energy-momentum tensor is quadratic as opposed to linear, there is a
contribution from the bulk energy-momentum tensor and there is a
contribution from the bulk Weyl tensor (note that no such term is
present in the five-dimensional Einstein equation).

The interpretation of the differences is straightforward. First,
it is by dimensional analysis clear that the relationship between the
induced Einstein tensor and the brane energy-momentum tensor cannot be
linear. The Einstein tensor has the dimension $m^2$, the
four-dimensional energy-momentum tensor $m^4$ and the
five-dimensional gravitational coupling $m^{-3}$. The first
combination of integer powers of the energy-momentum tensor and
gravitational coupling that has the same dimension as the Einstein tensor
is (e-m tensor)$^2$/(coupling)$^2$, the second is
(e-m tensor)$^5$/(coupling)$^6$ and so on. Without analysing the
five-dimensional Einstein equation and the Israel junction
conditions that determine the embedding of the brane into the
five-dimensional spacetime it is not obvious which possibility is
actually realised. However, it is clear that obtaining 
the standard linear dependence is impossible. In a Kaluza-Klein setting
where one integrates along the extra dimension, the ordinary
Einstein equation emerges (as an approximation) because the
size of the extra dimension provides a new dimensionful parameter.
However, in a brane setting the metric in the vicinity of a brane is
not sensitive to a global parameter such as the size of the extra
dimension, and so it does not appear in the induced Einstein equation.

The terms proportional to the bulk energy-momentum tensor simply account
for the fact that sources not confined to the brane can affect the
geometry on the brane. On the other hand, the term related to the bulk Weyl
tensor is somewhat surprising. The bulk Weyl tensor cannot be
solved from the local matter contribution, only from the complete
solution of the (five-dimensional) Einstein equation, and in this
sense it may be called non-local. Note that $E_{\mu\nu}$ is the
\emph{only} non-local term in the induced equation, and thus the
only one that may contain information about the global structure of
the five-dimensional spacetime. All information that an observer
localised on the brane can gravitationally obtain about the global
structure of the fifth dimension, such as its possibly finite size or
the presence of other branes, is included in $E_{\mu\nu}$.

\subsection{Gravity in the Randall-Sundrum model} \label{rsgravity}

In the Randall-Sundrum model, the bulk contains only a cosmological
constant $\Lambda$ and the visible brane has tension $\Lambda_1$ in addition
to its matter content. Then the induced equation \re{inducedgen} reduces to
\bea \label{inducedrs}
  ^{(4)} G_{\mu\nu} &=& \frac{\Lambda_1}{6 M_5^6} {}^{(4)}T_{\mu\nu} - \frac{1}{12 M_5^3}\left( 6 \Lambda + \frac{\Lambda_1^2}{M_5^3} \right) {}^{(4)}g_{\mu\nu} \el
  && + \frac{1}{24 M_5^6} \bigg( -6 ^{(4)}T_{\mu\alpha}\ ^{(4)}T^{\alpha}_{\ \nu} + 2 ^{(4)}T^{\alpha}_{\ \alpha}{} ^{(4)}T_{\mu\nu} \el
  && + \left( 3 ^{(4)}T_{\alpha\beta}\ ^{(4)}T^{\alpha\beta} -{} ^{(4)}T^{\alpha}_{\ \alpha}{} ^{(4)}T^{\beta}_{\ \beta} \right) {}^{(4)} g_{\mu\nu} \bigg) - E_{\mu\nu} \ ,
\eea

\noindent where $^{(4)}T_{\mu\nu}$ now stands for the energy-momentum tensor
of brane sources other than the tension, and the magnitudes of
$\Lambda$ and $\Lambda_1$ have been kept as free parameters not subject
to the fine-tuning \re{lambda}. The brane tension provides a local
dimensionful parameter that makes it possible to obtain a linear
dependence on the brane energy-momentum tensor. Remarkably, if the
contribution of the Weyl tensor is small and the scale of the energy
density is some orders of magnitude smaller than the five-dimensional
Planck scale, one obtains (nearly) standard four-dimensional gravity
on the brane.

Note that the flipside of possibly obtaining nearly standard gravity
with a positive brane tension even with a large extra dimension is
the impossibility of obtaining standard gravity without positive
brane tension, even with a small extra dimension\footnote{Assuming
that the bulk contains only a cosmological constant. It is clear from
\re{inducedgen} that it is also possible to obtain standard gravity by
putting an explicit dependence on the brane energy-momentum tensor in
the bulk energy-momentum tensor. For an example, see
\cite{Kanti:1999a, Kanti:1999b, Kanti:2000a, Kanti:2000b}.}.
This irrelevance of the size of the extra dimension to
the four-dimensional gravity is a distinctive feature of set-ups where
observers are localised on a brane, and contrasts sharply with
settings where observers are not localised in the extra dimensions.

From \re{inducedrs} we can identify the observed Newton's constant
and the observed cosmological constant as
\bea \label{constants}
  G_N \equiv \frac{1}{8\pi M_{Pl}^2} &=& \frac{\Lambda_1}{48\pi M_5^6} \el
  \Lambda_{eff} &=& \frac{3 M_5^3 \Lambda}{\Lambda_1} + \frac{\Lambda_1}{2} \ .
\eea

Two immediate observations from \re{constants} are that in order
to obtain a positive Newton's constant the brane tension has to be
positive, as mentioned earlier, and that in order to obtain a zero
effective cosmological constant on the brane there has to be a negative
bulk cosmological constant. The second point was already apparent in the
``vacuum'' set-up of Randall and Sundrum: in order to obtain a static
solution (which of course requires that the effective cosmological
constants on the branes vanish) the bulk cosmological constant had to be
negative and fine-tuned to both of the brane tensions, which is of course
only possible if they have the same magnitude.

Without the fine-tuning of the Randall-Sundrum proposal, the effective
cosmological constant can be adjusted to any desired value by choosing an
appropriate $\Lambda$. Of some interest is the case
$\Lambda=0$,\footnote{For $\Lambda=0$ there is of course no
exponential warping of spacetime.} since then the magnitude of the
effective cosmological constant is fixed in terms of the brane tension
and the five-dimensional Planck mass, or alternatively, the four- and
five-dimensional Planck masses. Putting $\Lambda=0$, the effective
cosmological constant is, from \re{constants},
\bea
  \Lambda_{eff} &=& \frac{3 M_5^6}{M_{Pl}^2} \ .
\eea

In order to obtain an effective cosmological constant in agreement
with observations
\cite{Perlmutter:1997, Riess:1998, Efstathiou:2001, Melchiorri:2002}
$\Lambda_{eff}\sim M_{Pl}^2 H_0^2\sim 10^{-48}$ (GeV)$^4$,
where $H_0$ is the present value of the Hubble parameter,
the five-dimensional Planck mass would need to be
$M_5\sim{} ^3\!\sqrt{H_0 M_{Pl}^2}\sim 10^{-2}$ GeV. However,
we will see in the next section that a value this small spoils
cosmology at the time of light element nucleosynthesis, quite apart from other
possible problems. So, the Randall-Sundrum model does not offer a
solution to the problem of a small non-zero cosmological constant.
However, there is hope of explaining a zero cosmological constant,
in other words the Randall-Sundrum fine-tuning, in a similar
setting in terms of supersymmetry \cite{Behrndt:1999, Brito:2001}.

While the effective cosmological constant, if different from zero, is in
principle observable, there is nothing that would distinguish it from
an ordinary four-dimensional cosmological constant. The second order terms
in the energy-momentum tensor are suppressed by
$\sim{}^{(4)}T_{\mu\nu} M_4^2/M_5^6$, so that unless the five-dimensional
Planck mass is very small, they will be quite difficult to observe. The
magnitude of the Weyl tensor-term $E_{\mu\nu}$ is not obviously suppressed, but
it cannot be evaluated without knowing the bulk solution. Brane
cosmology offers one clear way of assessing the observability of the
second order terms and minimises the uncertainty due to $E_{\mu\nu}$.

\subsection{Cosmology in the Randall-Sundrum model} \label{rscosmology}

Specialising the Einstein equation \re{inducedrs} to cosmology by
assuming homogeneity and isotropy with respect to the visible spatial
directions, we obtain \cite{Binetruy:1999b}
\bea \label{rshubble}
  & & \frac{\adot^2}{a^2} + \frac{K}{a^2} \,=\, \frac{\Lambda_1}{18 M_5^6}\rho + \frac{1}{36 M_5^3}\left( 6 \Lambda + \frac{\Lambda_1^2}{M_5^3} \right) + \frac{1}{36 M_5^6}\rho^2 + \frac{{\cal C}}{a^4} \el
  & & \rhodot + 3 \frac{\adot}{a} (\rho + p) \,=\, 0 \ ,
\eea

\noindent where $a$ is the scale factor of the brane, $6K$ is the
constant spatial curvature on the brane, $\rho$ and $p$ are the
energy density and pressure, respectively, of brane matter, ${\cal C}$
is a constant and a dot indicates a derivative with respect to proper
time on the brane.

The second equation is the ordinary four-dimensional conservation law.
A standard conservation law for brane matter is not a generic feature of
brane cosmology, but simply a consequence of the assumption that there is no
energy flow in the direction of the fifth dimension (since the bulk is empty
apart from the cosmological constant).

The first equation is the Hubble law. As discussed in the previous
section, the departures from standard cosmology are confined to the
$\rho^2$-term and the Weyl tensor contribution ${\cal C}/a^4$. As long
as the five-dimensional Planck scale $M_5$ is more than 10 TeV, the
effect of the $\rho^2$-term will be negligible from the time of
neutrino decoupling (at $\rho\sim1$ MeV) onwards. Conversely, the value
$M_5\sim 10^{-2}$ GeV that could naturally explain the present-day
cosmic acceleration is ruled out by the successful predictions of
standard big bang nucleosynthesis \cite{Olive:2002}. The magnitude of
the ${\cal C}$-term could only be known from a full bulk solution, but
it can in any case affect only the early universe since its
contribution declines faster than that of non-relativistic matter (for
which $\rho\propto a^{-3}$).

Like the general induced Einstein equation in the Randall-Sundrum
model, \re{inducedrs}, the Hubble law \re{rshubble} is not closed,
and cannot be solved without specifying ${\cal C}$, which depends on
the full bulk solution. Note that all possible
effects due to other branes or the finite size of the extra dimension
are contained in the value of the single constant ${\cal C}$. This feature,
while surprising from a Kaluza-Klein point of view, is quite natural in a
brane setting. As mentioned earlier, the five-dimensional Einstein equation
and the Israel junction conditions which give the embedding of the brane into
the five-dimensional spacetime are local, so that the four-dimensional Einstein
equation can contain a non-local contribution only via the Weyl tensor.
Since all contractions of the Weyl tensor with itself vanish, $E_{\mu\nu}$
is traceless, which means that its contribution to the Hubble law decays like
radiation. Therefore the only degree of freedom is the magnitude given
by the constant ${\cal C}$.\footnote{For this argument it is necessary that
the second order contribution of the brane energy-momentum tensor is
separately conserved. This is true in the homogeneous and isotropic case
but not in general.}

Apart from non-closure, the Hubble law \re{rshubble} has another shortcoming,
also shared by the general Einstein equation \re{inducedgen}: a solution
of the induced equation is not necessarily a solution of the full bulk
equation. In other words, the equation is a necessary but not a
sufficient condition for the solution to exist. In practical terms, this means
that even if one can solve the induced equation despite non-closure (for
example, by neglecting the ${\cal C}$-term at late times), one still has
to construct the bulk solution to know whether the cosmological
solution is actually realised. One can turn the issue upside down and
view this as a constraint imposed by the induced equation on the bulk
solutions. This point of view can be illustrated with the \emph{Israel
junction conditions} which determine the embedding of the branes into
the five-dimensional spacetime. For the isotropic and homogeneous
cosmological case with the metric
\bea
  ds^2 = - n(t,y)^2 dt^2 + \frac{a(t,y)^2}{(1+\frac{K}{4}r^2)^2} \sum^3_{j=1}(dx^j)^2 + b(t,y)^2 dy^2 \ ,
\eea

\noindent where $r^2=\sum^3_{j=1}(x^j)^2$ and the constant $6 K$ is the
spatial curvature in the directions parallel to the branes, the junction
conditions read \cite{Binetruy:1999a}
\bea \label{israel}
  \pm \frac{1}{b}\frac{a'}{a}\bigg|_{y=y_1}
&=& -\frac{1}{6 M_5^3} (\rho + \Lambda_1) \el
  \pm \frac{1}{b}\frac{n'}{n}\bigg|_{y=y_1}
&=& \frac{1}{6 M_5^3} (2\rho + 3 p - \Lambda_1) \ ,
\eea

\noindent where $y_1$ is the location of the brane, a prime denotes a
derivative with respect to $y$, and the sign ambiguity is related to whether
one takes the limit $y\rightarrow y_1$ from the right or from the left of
$y_1$ ($a'$ and $n'$ are discontinuous at the brane).

To demonstrate how the junction conditions relate the bulk metric to the
matter content of the brane, let us consider the following rather general
ansatz
\bea
  n(t,y) &=& n(t,y) \el
  a(t,y) &=& a(t) n(t,y) \el
  b(t,y) &=& b(t,y) \ ,
\eea

\noindent where $n(t,y)$, $a(t)$ and $b(t,y)$ are arbitrary functions.
Since $n'/n=a'/a$, the Israel junction conditions give the following 
constraint on brane matter
\bea
  \rho + p = 0 \ ,
\eea

\noindent which implies via \re{rshubble} that $\rhodot=0$: only vacuum
energy is allowed.

The above example shows that the four-dimensional part of the bulk metric
cannot be factorisable so that we would have
$g_{\mu\nu}(x^\mu,y)=f(y) ^{(4)}g_{\mu\nu}(x^\mu)$ or even
$g_{\mu\nu}(x^\mu,y)=f(x^\mu,y) ^{(4)}g_{\mu\nu}(x^\mu)$. In particular,
the Randall-Sundrum warp factor idea for solving the hierarchy problem is not
possible in a cosmological setting.

Given the constraints, one may wonder whether for an arbitrary solution
of the induced equation there exists a bulk solution that supports it.
The answer is in the affirmative. In the case of a single
brane, the explicit bulk solution for the case of a static fifth dimension,
$\bdot=0$, and arbitrary matter content $\rho(t)$ and $p(t)$ has been
constructed \cite{Binetruy:1999b}. The solution contains no free parameters
(apart from possibly ${\cal C}$), so that it is clear
that it will not be a solution if one includes a second brane with matter.
The situation has been studied, keeping $b$ constant but
allowing the position of the second brane to change in time
\cite{Binetruy:2001}. As expected, for general matter content on the two
branes there is no solution. The interpretation is presumably 
that two branes with matter on them will inevitably cause the size of the
fifth dimension to vary in time, not that solutions with matter on two
branes and only a cosmological constant in the bulk
do not exist\footnote{If one keeps the bulk energy-momentum tensor as set of
free parameters, solutions with a static fifth dimension may well exist. For
an example (where $b$ is time-independent to lowest order in a
perturbative expansion), see \cite{Kanti:2000a, Kanti:2000b}.}.
As pointed out in \cite{Enqvist:2000},
a similar result is obtained if one adds ideal fluid to the bulk:
there are no solutions with generic matter in the
bulk if one keeps the fifth dimension static.

Cosmology at the homogeneous and isotropic level gives little hope of
detectable signals of the Randall-Sundrum model. However, the
higher-dimensional setting of course modifies also the
equations that describe departures from homogeneity and isotropy.
Probably the most important among these are the perturbation equations that
govern the behaviour of the cosmic microwave background \cite{Perturbations},
which may offer a possibility for detectable signals.

\section{Summary}

Even though the Randall-Sundrum model and other brane set-ups thus far
presented do not solve the hierarchy problem, they offer an interesting new
approach to particle physics and cosmology. In particular, they have
challenged the old view of dimensional reduction via compactification
by providing an alternative which demonstrates that it is possible to
obtain nearly standard four-dimensional gravity with large or even
infinite extra dimensions.

The greatest shortcoming of most brane gravity models, including
the Randall-Sundrum model, is that they are not based on fundamental
principles. In most cases, the number of spacetime dimensions and
the brane structure are essentially unmotivated, as are the
contents of the bulk, be it a cosmological constant, a scalar field
or an ideal fluid. The confusion originating in uncertain foundations and
aggravated by the misapplication of old Kaluza-Klein
ideas\footnote{Such as trying to stabilise the extra dimension, which is not
only overly restrictive \cite{Enqvist:2000, Grinstein:2000, Binetruy:2001} but
also unnecessary, since the gravitational coupling constant on the
brane does not depend on the size of the extra dimension, unlike in
Kaluza-Klein settings.} is readily appreciated by sampling the
literature on Randall-Sundrum-type scenarios.

The ekpyrotic scenario is a realisation of the brane scenario that is
based on fundamental physics. Heterotic M-theory offers a brane set-up
grounded in a unified theory of gravity and particle physics
that motivates the dimension of spacetime (and the codimension of the
branes) and provides an explicit account of the contents of the bulk.
The application of cosmological brane ideas in heterotic M-theory
is is somewhat reminiscent of the application of the poorly motivated
Kaluza-Klein ideas in the well-defined arena of supergravity, with
the difference that the well-defined heterotic M-theory
existed already \emph{before} the Randall-Sundrum-inspired brane models.

While the origin of the ekpyrotic scenario is heterotic M-theory
and the framework is brane physics, the main motivation comes from
cosmology. Before proceeding to the ekpyrotic scenario, it is therefore
appropriate to review the status of cosmological scenarios that serve
as its backdrop.

\chapter{Scenarios of the primordial universe} \label{cosmo}

\section{Six cosmological problems}

The ekpyrotic scenario is based on heterotic M-theory and aims to give a
comprehensive description of the primordial universe. It was explicitly
presented as an alternative to the prominent scenarios of the primordial
universe, in particular inflation. I will therefore briefly review the
current cosmological problems, present the scenarios of the primordial
universe that have been proposed to address these problems and highlight
the shortcomings of these proposals.

The cornerstone of modern cosmology is the big bang theory. The
theoretical foundation of the big bang theory is very solid: one
simply applies general relativity to a homogeneous and isotropic
four-dimensional spacetime filled with matter that is treated as
an ideal fluid, taking into account atomic, nuclear and possibly
strong and electroweak physics in the early universe. The theory
is also in excellent agreement with observations. The main support
comes from the redshift of light emitted by distant objects, the
temperature of the cosmic microwave background (CMB) and the
abundance of the elements D, $^3$He, $^4$He and $^7$Li \cite{Olive:2002}.

However, the observational support for the big bang theory does not
extend to eras before the decoupling of neutrinos from nucleons and
electrons at about one second after the big bang. The only observables
we at present have from times before neutrino decoupling --which I
will refer to as the primordial era-- are the anisotropies of the
CMB
[71, 75-78],
large-scale structure \cite{Efstathiou:2000, Dodelson:2001}, baryon number,
the amount and properties of dark matter and possibly the
amount and properties of dark energy \cite{Turner:2001}. In addition, the
spatial curvature of the universe is an observable presumably related to
the primordial universe. None of the above observables have an
explanation in the context of the big bang theory (with the possible
exception of the baryon number via electroweak baryogenesis
\cite{Kuzmin:1985, Riotto:1999}).

It is also clear from a theoretical point of view that the big bang theory
is not the correct description of the primordial universe, since the high
energy densities and curvatures near the big bang imply breakdown of the
classical treatment of gravity on which the theory is based.

For a description of the primordial universe one has to turn to a
something beyond general relativity plus ideal fluid. It is to be
expected that this theory would explain not only the observables that
the big bang theory cannot account for, but also the starting point of
the big bang theory: the homogeneity and isotropy of the universe and
the origin of matter. Specifically, a theory of the primordial universe
should cover the following things.

\begin{enumerate}

\item The origin of matter

The origin and relative amount of both visible and dark matter in
the universe is a fundamental question confronting scenarios of the
primordial universe. In the big bang theory, matter is always present,
and there is no explanation as to the relative amount of visible
and dark matter, or matter and antimatter. The amount and properties
of dark energy, if it is not purely geometric, also falls
in the same category.

In considering the generation of matter, I will leave aside the question of
the nature of the matter generated. Also, I will not address the question of
dark energy, since no presently favoured scenario of the primordial
universe sheds any light on it.

\item The homogeneity and isotropy of the universe

The universe is spatially quite homogeneous and isotropic. In particular,
the temperature of the cosmic microwave background is the same in all
directions to an accuracy of $10^{-5}$. This is a puzzling observation,
since it means that parts of the universe that have never been in causal
contact according to the big bang theory have nevertheless almost exactly
the same conditions, in an apparent violation of locality.

\item The spatial flatness of the universe

The universe appears to be nearly spatially flat. Unlike homogeneity and
isotropy, this does not constitute a mystery or imply a violation of any
physical principles within the confines of the big bang theory. There are
three possibilities for the spatial geometry of a spatially homogeneous and
isotropic universe: open, closed and flat. One would like to explain via
some physical principle why the universe happens to possess one particular
spatial geometry out of the three possibilities, but no fine-tuning is
involved.

\item The seeds of large-scale structure

The problem of homogeneity and isotropy is somewhat vague due to the
scarcity of observables, and it is the departures from homogeneity and
isotropy that are the main quantifiable predictions for present scenarios of
the primordial universe. The data on the inhomogeneities comes from
observations of the cosmic microwave background
[71, 75-78]
and large-scale structure \cite{Efstathiou:2000, Dodelson:2001}.
In particular, any contender for a model of the
primordial universe should explain the origin and nature of the
anisotropies of the CMB with sufficient accuracy to be compared with the data.

The most important observationally confirmed aspects of the anisotropies of
the CMB are that they are mostly \emph{adiabatic} and their amplitude is
almost scale-independent and about $10^{-5}$. Precise definitions can be
found in the literature \cite{Linde:1990, Kolb:1994}. Roughly speaking,
adiabaticity means that the perturbations are along the same direction in
field space as the background (the alternative would be \emph{isocurvature}
perturbations), scale-invariance means that the amplitude
does not depend on the wavelength, and the amplitude is simply the maximum
relative difference between the perturbed cosmic microwave background
temperature and the average temperature, $(T_{max}-T_{av})/T_{av}$.

The scale-dependence of the perturbations is usually expressed with the
spectral index $n$. An index less than 1 means that the amplitude of large
wavelengths is amplified with respect to small wavelengths, leading to a
``red spectrum'', an index of 1 means that all wavelengths have the same
amplitude, leading to a scale-invariant spectrum, and an index of more than
1 means that small wavelengths are amplified with respect to large ones,
leading to a ``blue spectrum''. According to observations,
$n=1.03^{+.10}_{-.09}$ \cite{Netterfield:2001}.

\item The absence of topological defects

Phase transitions at high energies in grand unified theories of particle
physics are expected to produce monopoles, cosmic strings and domain walls.
According to the big bang theory, the energy density of the primordial
universe is high enough to produce an abundance of such relics, yet none are
observed. Unlike the previous problems, the defect problem is one of
non-observation, and therefore more vague. While the topological defect
problem, also known as the relic problem, has been treated as a shortcoming
of cosmology, it is not impossible that the issue might be resolved in the
realm of particle physics instead.

\item The singularity problem

Unlike the previous five problems, which were observational (albeit, in the
case of the topological defect problem, in a negative sense), the
singularity problem is purely theoretical. According to the big bang theory,
the universe began in a state of infinite curvature and energy density a
finite time ago. While singularities as such may not necessarily be
unphysical, it is clear that unbounded curvatures and energy densities imply
that the classical treatment of gravity and matter cannot be trusted.

A consistent model of the primordial universe would either need to be
non-singular or have a singularity that does not lie outside its domain of
validity. An example of the latter would be an initial singularity which is
not a curvature singularity and does not involve unbounded energy densities.
A more modest requirement is to have a model where the singularity is
observationally and theoretically irrelevant in the sense that the
observables are not sensitive to the singularity and the internal
consistency of the model is not degraded by the singularity. The big bang
theory, for example, satisfies both of these requirements.

\end{enumerate}

At the moment, there are two prominent scenarios of the primordial
universe, the inflationary scenario and the pre-big bang scenario.
I will now briefly review these scenarios and assess how they
address the abovementioned problems.

\section{The inflationary scenario}

The currently favoured framework for addressing primordial cosmological
problems is inflation. Inflation is not a firmly established theory but
rather a scenario which finds its realisation in a number of different
models. The scenario posits that a patch of the primordial universe started
to undergo accelerating expansion, \emph{inflation,} and that the presently
observable universe originates from a small (in many models trans-Planck
scale) volume of the primordial universe. In many models the universe starts
with a big bang, after which the universe is supposed to be in an unordered
state of high energy density and curvature, from which the inflationary
patch emerges. For reviews on inflation, see
\cite{Brandenberger:1999, Linde:2002}.

The set-up for most models of inflation is big bang theory modified in a
simple way, by just adding a causative agent for inflation.
In most models the source of inflation is the potential energy of one
or more scalar fields. The scalar field(s) evolve slowly, so that the
expansion lasts long (in units of the potential energy, which varies between
different models, but is typically a few orders of magnitude below
the Planck scale). There are other possibilities for the source of
expansion, most notably vacuum energy \cite{Tsamis:1996, Abramo:1998}
and higher order curvature terms due to quantum fields
\cite{Starobinsky:1980, Starobinsky:1983}.
The action for most scalar models can be written as
\bea \label{scalaraction}
  S_{\textrm{scalar}} \,=\, \frac{1}{16\pi G_N}\int d^4 x \sqrt{-g}\left(R -
\sum_{i=1}^n\frac{1}{2}\pat_{\mu}\phi^i\,\pat^{\mu}\phi^i - V(\phi^i)\right) \ ,
\eea

\noindent where $G_N$ is Newton's constant, $R$ is the scalar curvature,
$\phi^i$ are some scalar fields and $V(\phi^i)$ consists of mass and
interaction terms. Typical examples are chaotic inflation
\cite{Linde:1983} with a single scalar field and $V(\phi)=\frac{1}{2}m^2
\phi^2$ or $V(\phi)=\frac{1}{4}\lambda\phi^4$ and hybrid
inflation \cite{Linde:1993} with two scalar fields and
$V(\phi, \chi)=\frac{1}{2} m^2 \phi^2 + \frac{1}{2}\lambda' \phi^2 \chi^2 + \frac{1}{4}\lambda(M^2-\chi^2)^2$.

The action for the vacuum energy model is \cite{Tsamis:1996, Abramo:1998}
\bea \label{vacuumaction}
  S_{\textrm{vacuum}} \,=\, \frac{1}{16\pi G_N}\int d^4 x \sqrt{-g}\left(R - 
\Lambda \right) \ ,
\eea

\noindent where $\Lambda$ is a cosmological constant. Treating
general relativity as a quantum field theory makes the model considerably
more complicated than the simplicity of the action would seem to indicate.
(Counterterms needed to absorb the divergences of the quantised theory have
been omitted from \re{vacuumaction}.)

A typical action for a higher order curvature model is \cite{Starobinsky:1980, Starobinsky:1983}
\bea \label{curvatureaction}
  S_{\textrm{curvature}} \,=\, \frac{1}{16\pi G_N}\int d^4 x \sqrt{-g}\left(R
+ a R^2 + b R_{\alpha\beta}R^{\alpha\beta} + c R_{\alpha\beta\gamma\delta} 
R^{\alpha\beta\gamma\delta} \right) \ ,
\eea

\noindent where $a,b$ and $c$ are constants, $R_{\alpha\beta}$ is
the Ricci tensor and $R_{\alpha\beta\gamma\delta}$ is the Riemann
tensor. (The model outlined in \cite{Starobinsky:1980, Starobinsky:1983}
also includes contributions that cannot be expressed in terms of a
local action.)

Whatever the mechanism, in all models of inflation the universe expands by a
huge factor, typically more than $e^{60}$, during the primordial era. The
accelerating expansion eliminates almost all traces of the conditions before
inflation. The six cosmological problems are addressed in the following
manner.

\begin{enumerate}

\item The origin of matter

In the inflationary scenario, the origin of possible matter in the
primordial universe is not explained and is largely irrelevant, since it is
diluted by the large expansion. The relevant question is the creation (and
thermalisation) of the matter observed today, so-called ``reheating'' or
``preheating''. This problem is not yet entirely solved. The main proposed
mechanisms are particle production due to a scalar field oscillating about
the bottom of its potential \cite{Traschen:1990, Kofman:1997} and
gravitational particle production in expanding spacetime \cite{Ford:1987, Felder:1999}.

\item The homogeneity and isotropy of the universe

Homogeneity and isotropy are supposed to be explained by the huge stretching
of spacetime: any inhomogeneities and anisotropies are diluted by the
expansion. However, it has been shown \cite{Trodden:1998} that assuming the
weak energy condition\footnote{The weak energy condition states that
$T_{\mu\nu}u^{\mu} u^{\nu}\ge0$ for all timelike vectors $u^{\mu}$.}
it is impossible to start inflation unless there is
already isotropy and homogeneity on scales larger than the causal horizon.
The weak energy condition is satisfied in all scalar field models with
standard kinetic terms, as well as in the vacuum energy model, though it can
be broken by higher order curvature terms. So, at least the scalar field and
vacuum models of inflation can only ameliorate the homogeneity and isotropy
problems, not solve them, though that is sometimes mentioned as their main
motivation. The problem can be minimised by starting inflation as near the
Planck scale as possible, so that the coincidence of initial homogeneity
and isotropy is hopefully as small as possible.

\item The spatial flatness of the universe

Spatial flatness is explained in the same way as homogeneity and isotropy:
any possible spatial curvature of the universe is stretched to an
unobservably small level due to the vast expansion, so that the universe
looks spatially flat on presently observable scales.

\item The seeds of large-scale structure

The origin of the large-scale structure of the universe is thought to lie in
quantum fluctuations  -- in the scalar field case, the fluctuations of one
or more scalar fields, in the vacuum energy case the fluctuations of the
quantised spacetime metric, and in the higher order curvature term case in
the spacetime curvature. These fluctuations leave a definite imprint on the
homogeneous and isotropic background and grow to become the CMB anisotropies
and the seeds of galaxies. The prediction of the definite type (mostly
adiabatic) and shape (almost scale-invariant) of the CMB anisotropies is at
present the most solid support of the inflationary scenario, particularly
the simplest scalar field models. However, in these simplest scalar field
models the amplitude of the perturbations is not naturally explained, since
one has to generically tune some parameters to be quite small to obtain a
small enough amplitude \cite{Adams:1991}. For example, for chaotic inflation
with the potential $V(\phi)=\frac{1}{4}\lambda\phi^4$, one has to take
$\lambda\sim 10^{-12}$ \cite{Brandenberger:1999}.

\item The absence of topological defects

The monopoles, cosmic strings and other possible relics of the
primordial high-curvature era are diluted to unobservable densities by
inflation. One might then hope that if the reheating temperature is low
enough, the relics would not reappear. However, it is possible to produce
too many relics at reheating even if the temperature is several orders of
magnitude lower than the energy scale at which the unwanted objects are
typically produced \cite{Felder:1999}. So, the problem is not completely
solved.

\item The singularity problem

It has been shown that any manifold on which the local Hubble parameter
measured by an observer on a null or timelike geodesic is bounded from below
by a positive constant in the past is singular \cite{Borde:2001}. No
symmetry arguments or energy conditions are required, and most remarkably,
the proof does not make use of the Einstein equation. This theorem implies
that all scalar field models and the vacuum energy model as well as many
higher order curvature models are singular. Furthermore, the modest
requirements of observational and theoretical irrelevance are not quite
satisfied, so that the singularity problem is in a sense worse than in the
big bang theory. Since observables such as the CMB anisotropies are produced
in the latter stages of inflation, they are insensitive to the initial
singularity, satisfying the requirement of observational irrelevance.
However, the requirement of theoretical irrelevance is not satisfied.

It is preferable to start inflation as near the singularity as possible
in order to have to postulate as little homogeneity and isotropy as
possible, as noted in connection with the homogeneity and isotropy problem.
But one would have to know the distribution of fields and curvature near the
singularity in order to evaluate how probable it is to have the desired
homogeneous and isotropic volume larger than the causally connected volume.
If the universe begins in a curvature singularity, a well-defined initial
value problem that would allow for a rigorous treatment of this problem of
course cannot be formulated. If the curvature is bounded at the singularity, it
may be possible to formulate the initial value problem given the boundary
conditions, for example by an instanton describing the birth of the universe
\cite{Vilenkin:1985}. However, the issue is complicated by higher order
curvature terms, discussed below.

\end{enumerate}

In addition to the open problems mentioned above, most prominently starting
inflation and treating the singularity, there are two distinct but related
problems which bear mentioning.

First, most models of inflation (the vacuum energy model being an
interesting exception) are semiclassical theories of quantum gravity: the
scalar fields (and the fluctuation modes of the metric) are treated
quantum mechanically, while the background geometry is classical. In
semiclassical quantum gravity, quantum fields induce higher orders of
curvature such as those in \re{curvatureaction} into the action
\cite{Birrell:1982}. The contribution of such terms is typically suppressed
by the Planck mass, so that one might be tempted to argue that their effect
is small, apart from considerations of the initial value problem as
discussed above. However, since such terms generally contain fourth and
higher order time derivatives, they can completely change the behaviour of
the equations of motion, regardless of how small their coefficients are
\cite{Horowitz:1978}. The situation is aggravated by the fact that inflation
typically lasts long, allowing ample time for gravitational instabilities to
develop. Usually such terms are simply ignored, but it should be understood
that it is not mathematically consistent to do so, nor necessarily
physically justified\footnote{For a mathematically consistent but not
necessarily physically justified treatment of the higher order terms,
relevant also for higher order curvature inflation, see
[104-106].}.

Second, in many models of inflation the perturbations that seed the
large-scale structure seen today originate from physical wavelengths many
orders of magnitude smaller than the Planck length. It is questionable whether
one can apply semiclassical quantum field theory of free fields in the case of
wavenumbers and frequencies many orders of magnitude larger than the Planck
mass; even less applicable is the linear perturbation theory that is used to
calculate the behaviour of the perturbations. In the light of string theory
and other theories of quantum gravity \cite{Carlip:2001}, it may not make any
sense at all to speak of distances shorter than the Planck length.

\section{The pre-big bang scenario}

The main alternative to inflation as a comprehensive scenario of the
primordial universe is the pre-big bang scenario. As noted above, the
approach in most models of inflation is to take general relativity and add
an inflation-producing agent by hand, with the expectation that the set-up
may later find justification in a more fundamental framework. There is
usually no connection to a unified theory of gravity and particle physics,
or to a theory of quantum gravity (the vacuum energy model being a
remarkable exception). The pre-big bang scenario, like the ekpyrotic
scenario, takes the opposite approach and attempts to descend from a
promising unified theory of quantum gravity, namely string theory, down to
phenomenology. For recent reviews of the pre-big bang scenario, see
[108-110].

The framework of the pre-big bang scenario is ten-dimensional superstring
theory compactified down to four dimensions. The simplest effective
four-dimensional action usually considered is
\bea \label{PBBaction}
  S_{\textrm{PBB}} \,=\, \frac{1}{16\pi G_N}\int d^4 x \sqrt{-g}\left(R - \frac{1}{2}\pat_{\mu}\phi\,\pat^{\mu}\phi - \frac{1}{2}\pat_{\mu}\beta\,\pat^{\mu}\beta - \frac{1}{2}e^{2\phi}\pat_{\mu}\sigma\,\pat^{\mu}\sigma\right) \ ,
\eea

\noindent where the scalar fields $\phi$, $\beta$ and $\sigma$ are the
dilaton, modulus and axion, respectively.

The four-dimensional universe described by the above action is supposed to
start in the far past in a ``trivial'' state with low curvature and energy
density. Due to an instability the universe starts collapsing and the
curvature starts growing. The universe is supposed to ``gracefully exit''
(due to terms not present in the above action) from the collapsing phase to
the usual expanding phase before reaching the ``big crunch'' singularity.
The collapse in the pre-big bang scenario plays a role similar to expansion
in the inflationary scenario. The six cosmological problems are addressed in
the following manner.

\begin{enumerate}

\item The origin of matter

The pre-big bang scenario is not yet definite enough to give an account of
the generation of matter. However, matter production is expected to occur
during the graceful exit phase, and may even play an important role in
achieving the reversal of contraction to expansion.

\item The homogeneity and isotropy of the universe

The homogeneity and isotropy of the universe is provided by a period of
accelerating collapse which makes the universe homogeneous and isotropic
in the same way as accelerating expansion. (Indeed, in the conformally
related metric known as the ``string frame'', which is as physically relevant
as the usual Einstein metric, the collapse looks like accelerating
expansion.)

\item The spatial flatness of the universe

During the collapsing phase, spatial curvature grows instead of decreasing
as in inflation. However, the growth is slower than the growth of the energy
density of the dilaton field, so that the contribution of spatial curvature
to the dynamics of the universe becomes negligible. (In the string frame the
increasing spatial flatness is more transparent: the accelerating expansion
dilutes spatial curvature.)

\item The seeds of structure

As in the inflationary scenario, the seeds of large-scale structure are
quantum fluctuations about an isotropic and homogeneous background. In the
original pre-big bang set-up, the fluctuations were those of the dilaton
field which is the source for the collapse. They are adiabatic, since
they are fluctuations of the quantity driving the background evolution.
However, since the Hubble parameter is not constant but rapidly decreasing, the
fluctuations are not scale-invariant but deeply blue, with $n\approx4$. This
problem was solved by considering the fluctuations of another field, the
axion. It is possible to obtain nearly scale-invariant fluctuations for the
axion field, but since it does not contribute to the background dynamics,
these fluctuations will be of the isocurvature type instead of
adiabatic. On a positive note, the amplitude of the fluctuations is set by
the string scale and the spectral index, and it is possible to obtain the
correct amplitude without the introduction of new small parameters.

There have recently been two proposals to obtain a spectrum of
scale-invariant adiabatic fluctuations in agreement with observation. The idea
of the first proposal \cite{Enqvist:2001b, Lyth:2001b} is that in the
post-big bang era the axion field oscillates about a minimum, and eventually
comes to dominate the energy density of the universe. The field then decays
into photons, so that the nearly scale-invariant isocurvature fluctuations of
the axion field are converted into nearly scale-invariant adiabatic
fluctuations of the photon background. In this proposal, the amplitude of
the fluctuations is set by the potential of the axion field, much as in
inflation the amplitude is set by the potential of the scalar field driving
inflation. However, unlike in inflation, no very small parameters are needed
and it is possible to naturally obtain the correct amplitude of density
perturbations.

In the second proposal \cite{Finelli:2001} one adds an exponential
potential for the dilaton, plus a new scalar field with a non-minimal
coupling to gravity. With both of these fields contributing to the background,
one can apparently obtain a scale-invariant spectrum of adiabatic
fluctuations for the dilaton field, and possibly also for the axion field.
However, the identification of the modes of the pre- and post-big bang
phases has been criticised \cite{Hwang:2002}.

\item The absence of topological defects

Since there is no exponential stretching of spacetime after the big bang,
there seems to be no mechanism for diluting the abundance of dangerous relics.
The problem might possibly be solved by having a low enough energy density
at the big bang.

\item The singularity problem

In the pre-big bang scenario the collapsing era during which the
anisotropies of the CMB are produced and the current expanding era are
separated by the big crunch/big bang curvature singularities. In this
context, the singularity problem appears in the guise of joining these two
eras in a non-singular manner, known as the \emph{graceful exit}.

On the one hand, the singularity problem in the pre-big bang scenario is more
severe than in the big bang theory, or in the inflationary scenario, since the
curvature singularity does not occur before but between eras of cosmology that
produce observables.

On the other hand, the pre-big bang scenario offers a solid framework for
avoiding the singularity, unlike scalar field models of inflation. String and
quantum corrections can provide various modifications to the equations of
motion that might resolve the singularity. The work thus far seems to indicate
that the graceful exit, if it can be achieved at all, is likely to occur in the
strongly coupled regime of string theory 
[115-118]. Given that little is known about strongly coupled
string theory, the assumption that the perturbations generated during the
pre-big bang era can be transferred to the post-big bang era with simple
matching conditions which are insensitive to the physics of the graceful exit
seems questionable. It does not seem impossible that the largely
unknown physics of the exit era could change the perturbations radically.
Until this problem has been solved, all predictions of the pre-big bang
scenario can only be considered preliminary.

\end{enumerate}

In addition to the problems mentioned above, one could mention that the
issue of initial conditions is not settled. The pre-big bang scenario does
not suffer from a singularity problem in the era before graceful exit, so
that a formulation of the initial value problem is possible, in contrast to
scalar field inflation. The issue of initial conditions is related to the
duration of collapse. Since the magnitude of the Hubble parameter increases
during the collapse and the graceful exit should come into play before the
Hubble parameter exceeds the string scale, its initial value provides a
bound on the amount of collapse (or, in the string frame, expansion)
possible. At the moment, this and other issues related to the initial
conditions \cite{Kaloper:1998} remain open.

\section{The ekpyrotic scenario}

The next chapter will deal with the ekpyrotic scenario in detail, but
I will here give a brief qualitative account of the scenario and outline
the solutions it offers to the six cosmological problems.

Like the pre-big bang scenario, the ekpyrotic scenario starts from a
fundamental, though speculative, unified theory. As mentioned in chapter
\ref{rs}, the starting point of the ekpyrotic scenario is five-dimensional
heterotic M-theory, where the fifth dimension terminates at two boundary
branes, one of which is identified with the visible universe. There are two
different versions of the ekpyrotic scenario, the ``old scenario'', where
there is a bulk brane between the boundary branes and the ``new scenario'',
where only the boundary branes are present.

In both scenarios the initial state is supposed to be very near the vacuum
state, where the branes are flat, parallel and empty. The vacuum is of
course static, and dynamics follow from a small breaking of the
supersymmetry, in the form of a very weak potential for interbrane distance.
The potential is taken to be attractive so that it draws the branes --in the
old scenario the bulk brane and the visible brane, in the new scenario the
boundary branes-- towards each other until they collide, an event called
\emph{ekpyrosis}. In the old ekpyrotic scenario, the bulk brane is absorbed
into the visible brane in a \emph{small instanton phase transition}, while
in the new scenario the boundary branes bounce apart after the
collision. Ekpyrosis is the defining feature of the ekpyrotic scenario, and
most of the cosmological problems are explained in terms of this collision
or in terms of symmetries related to the branes as follows.

\begin{enumerate}

\item The origin of matter

A significant fraction of the kinetic energy of the moving brane is supposed
to be converted into a thermal bath of radiation on the visible brane,
providing the matter content of the universe.

\item The homogeneity and isotropy of the universe

Because the branes are (almost) parallel they will collide at (almost)
the same time at all their points, producing an energy density with
an (almost) constant temperature, ``ekpyrotic temperature'',
everywhere in the visible universe.

\item The spatial flatness of the universe

Spatial flatness of the visible universe follows from the assumption of
starting very near the vacuum, where the branes are flat.

\item The seeds of large-scale structure

Though the branes start flat and parallel, they undergo quantum fluctuations
during their journey across the fifth dimension. Due to these ``brane
ripples'', some parts of the branes collide somewhat earlier or later than
the average, resulting in slightly cooler or hotter regions, respectively.
These primordial perturbations then grow to become the cosmic microwave
background anisotropies and seed the large-scale structure seen in the
universe today. The adiabaticity of the perturbations is easy to understand
in the formalism where the interbrane distance appears as a scalar
field\footnote{This formalism will be considered in more detail in chapter
\ref{ekpyrotic}.}: then the perturbations are obviously in the same
direction in field space as the background. The spectral index is supposed
to be nearly scale-invariant since the conditions change very slowly during
the journey of the brane(s). In the formalism there are a number of free
parameters whose natural magnitude it is difficult to estimate, so that it is
not clear how natural it is to obtain the correct small amplitude for the
perturbations, but at least it is easy, for the same reason.

\item The absence of topological defects

The production of unwanted relics is highly suppressed if the ekpyrotic
temperature is lower than the energy scale at which such relics are produced. 
It should be noted that, in contrast to the inflationary scenario, the
temperature of the universe is at no time higher than the ekpyrotic
temperature.

\item The singularity problem

In the ekpyrotic scenario the big bang is ignited at some finite temperature
and there are no curvature singularities. Since the scenario is based on
heterotic M-theory, the singularity theorems of general relativity, which
is a low-energy approximation of M-theory, do not
necessarily apply. However, the ekpyrotic scenario
does not include a description of what happened before the start of brane
movement. Any model where time does not extend infinitely far into the past
(and that does not contain closed timelike curves) is of course geodesically
incomplete and thus singular.

\end{enumerate}

In addition to solving cosmological problems, the ekpyrotic scenario also
proposes to solve problems of particle physics. The small instanton phase
transition in the brane collision may change the instanton number of the
visible brane and break the gauge group from $E_8$ to some smaller group,
for example $SU(3)\times SU(2)\times U(1)$ or $SU(5)$. It can also set the
number of light families to three. These are interesting directions, but
quite different from the cosmological issues, so they will not be discussed
here further.

The initial conditions of the ekpyrotic scenario, like those of the
inflationary scenario and the pre-big bang scenario, have been under debate.
The initial conditions has been criticised for fine-tuning, since the
dynamical evolution in the ekpyrotic scenario must start extremely near the
vacuum state \cite{Kallosh:2001a}. Starting the dynamical evolution nearly
but not quite in some special symmetric state does seem
unappealing. However, criticism along these lines is not terribly fruitful.
First, we will never (or at least not in the foreseeable future) be able to
measure the initial state of the universe. Second, the naturalness of the
size of some parameters in a given theory cannot be properly assessed until
it is known how these parameters arise. Thus far the symmetry breaking has
been added by hand, and it seems premature to conclude anything one way or
another until it has been actually derived from heterotic M-theory. (A
similar argument could be fielded in defense of the parameters responsible
for the amplitude of CMB perturbations in scalar field models of inflation.)

As an aside, let us note that the ``cyclic model of the universe''
[13-15]
was in part motivated by a desire to obtain the highly symmetric
initial conditions as a result of a dynamical process.

The ekpyrotic scenario has a number of problems. They will be considered in
the next chapter after a more detailed account of how the scenario is
supposed to work.

\section{Summary}

There are a few promising scenarios of the primordial universe and several
well-studied models that at least partially realise these scenarios. However,
at the present time there is no model that would give satisfactory
answers to all six cosmological problems outlined. Also, all current models
have some deep unsolved problems which are not merely technical. As noted,
the ekpyrotic scenario was presented as an alternative to the inflationary
and pre-big bang scenarios, in part motivated by these problems. The last
chapter is devoted to a more detailed account of the ekpyrotic scenario and
the problems that it in turn faces.

\chapter{The ekpyrotic scenario} \label{ekpyrotic}

\section{The set-up}

As mentioned in chapter \ref{cosmo}, there are two versions of the
ekpyrotic scenario: the old scenario, where there is a third brane in
the bulk, and the new scenario, where there are only the boundary branes.
In addition, there is a spin-off, the so-called ``cyclic model of the
universe'' 
[13-15],
which shares many
features with the ekpyrotic scenario. This chapter will be devoted
to a review of the old and new ekpyrotic scenarios and their shortcomings,
with some comments on the cyclic model at the end, in section \ref{cyclic}.

As mentioned in chapter \ref{intro}, the ekpyrotic scenario is
based on heterotic M-theory. The action for both the old and the new
version of the ekpyrotic scenario consists of three parts:
\bea \label{action}
  S = S_{het} + S_{BI} + S_{matter} \ ,
\eea

\noindent where $S_{het}$ is the action of five-dimensional heterotic
M-theory with minimal field content, $S_{BI}$ describes the brane interaction
responsible for brane movement and $S_{matter}$ describes brane matter
created in the brane collision.

\paragraph{Simplified heterotic M-theory.}

Five-dimensional heterotic M-theory contains a large number of fields
\cite{Lukas:1998a, Lukas:1998b}, so that dynamical analysis is quite difficult.
In the ekpyrotic scenario a pruned version of the theory is obtained by
considering the minimal field content, that is, by putting to zero all
fields whose equation of motion allows it. The resulting simplified action
is \cite{Khoury:2001a, Kallosh:2001b, Lukas:1998a, Lukas:1998b}
\bea \label{hetaction}
  S_{\textrm{het}} \ &=& \frac{M_5^3}{2}\int_{{\cal M}_5} d^5 x \sqrt{-g}\left(R-\frac{1}{2}\pat_A\phi\,\pat^A\phi-\frac{3}{2}\frac{1}{5!}e^{2\phi}{\cal F}_{ABCDE}{\cal F}^{ABCDE}\right) \el
  & & - \sum_{i=1}^3 3\alpha_i M_5^3\int_{{\cal M}_4^{(i)}}\!\! d^4\xi_{(i)}\bigg(\sqrt{-h_{(i)}}e^{-\phi} \el 
  & & - \frac{1}{4!}\epsilon^{\mu\nu\kappa\lambda}{\cal A}_{ABCD}\pat_{\mu}X^A_{(i)}\pat_{\nu}X^B_{(i)}\pat_{\kappa}X^C_{(i)}\pat_{\lambda}X^D_{(i)}\bigg) \ ,
\eea

\noindent where $M_5$ is the Planck mass in five dimensions, $R$ is the
scalar curvature in five dimensions, $e^{\phi}$ is the ``breathing
modulus'', which describes the size of the Calabi-Yau threefold and
${\cal A}_{ABCD}$ is a four-form gauge field, with field strength
${\cal F}=d{\cal A}$. The Latin indices run from 0 to 4 and the Greek indices
run from 0 to 3. The four-dimensional manifolds ${\cal M}_4^{(i)}$, $i=1,2,3$,
are the visible, hidden and bulk branes respectively, with internal coordinates
$\xi^{\mu}_{(i)}$ and tensions $3 \alpha_i M_5^3 e^{-\phi}$. Note that the
brane tensions can vary in time (and space) since they depend on the
breathing modulus. The coefficients $\alpha_i$ have to sum to zero
\cite{Donagi:2001}, and are parametrised as $\alpha_1=-\alpha$,
$\alpha_2=\alpha-\beta$ and $\alpha_3=\beta$, with $\beta>0$. The tensor
$g_{AB}$ is the metric on ${\cal M}_5$ and $h^{(i)}_{\mu\nu}$ are the
induced metrics on ${\cal M}_4^{(i)}$. The functions
$X^A_{(i)}(\xi_{(i)}^{\mu})$ are the coordinates in ${\cal M}_5$ of a point
on ${\cal M}_4^{(i)}$ with coordinates $\xi_{(i)}^{\mu}$, in other words
they give the embedding of the branes into the five-dimensional spacetime.

The ``vacuum'' of the above action is a BPS state, which is
invariant under Poincar\'e transformations in the directions parallel
to the brane and preserves one-half of the eight supersymmetries of
the five-dimensional theory. In the vacuum state the branes are flat,
parallel and static, so that their embedding is simply given by
($t\equiv x^0$, $y\equiv x^4$)
\bea \label{embed}
  X^A_{(i)}(\xi_{(i)}^{\mu}) = (t, x^1, x^2, x^3, y_i) \ ,
\eea

\noindent with $y_1=0$, $y_2=R$ and $y_3=Y$, where $R$ and $0<Y<R$ are
constants. 
The metric and the fields in the vacuum state are given by
\bea \label{BPS}
  ds^2 &=& -N^2 D(y) dt^2 + A^2 D(y)\sum^3_{j=1}(dx^j)^2 + B^2 D(y)^4 dy^2  \el
  e^{\phi(y)} &=& B D(y)^3 \el 
  {\cal F}_{0123y}(y) &=& -\alpha A^3 N B^{-1} D(y)^{-2} \qquad\qquad y\le Y \el
  & & - ( \alpha-\beta ) A^3 N B^{-1} D(y)^{-2} \ \quad y\ge Y \ ,
\eea

\noindent where $D(y)=\alpha y - \beta(y-Y)\theta(y-Y)+C$ and $N, A, B$ and
$C$ are constants. In the new ekpyrotic scenario, where there is no bulk
brane, the above equations hold with $\beta=0$.

\paragraph{Brane interaction.}

The vacuum is of course static. Dynamics are provided by a small breaking of
the supersymmetry in the form of non-perturbative M-theory interactions
between the branes, mediated by the exchange of M2-branes. In principle it
should be possible to obtain the resulting effective potential for the brane
distances from M-theory, but so far a potential with the desired properties
has not been derived. At present the potential has been added by hand to a
four-dimensional effective theory and even an effective treatment of the
brane interaction in five dimensions is lacking.

However, whatever the genesis or the exact form of the interaction, it
presumably approaches zero as the boundary branes approach each other, at
least in the new ekpyrotic scenario. The ``fifth dimension'' is the eleventh
dimension of the full theory, so that its size is given by the value of the
dilaton, in other words the string coupling constant\footnote{The matter may
not be so simple. The Newton's constant on the brane is not given by the
usual heterotic result \cite{Enqvist:2001a}, so that it is not clear whether
the usual identifications for the other constants are correct either. The
issue is discussed in section \ref{stab}.}. As the size goes to zero, the
coupling constant vanishes. The vanishing of the interaction between the
bulk brane and the visible brane at their collision, which is assumed in the
old scenario, is less obvious.

The brane interaction is supposed to be attractive, so that in the old
scenario the bulk brane is attracted to the visible brane, while
in the new scenario the boundary branes are attracted to each other.
The branes are assumed to remain flat and parallel (apart from quantum
fluctuations), so that in the old ekpyrotic scenario their embedding
differs from that of the BPS state \re{embed} only via the
time-dependence of $Y$. In the new scenario the embedding is the same
as in the BPS state (apart from quantum fluctuations).

The spatial homogeneity and isotropy in the directions parallel to the
branes is also assumed to be maintained during their journey (again,
apart from quantum fluctuations). The metric can be without loss of
generality written as
\bea \label{metric}
  ds^2 = -n(t,y)^2 dt^2 + a(t,y)^2 \sum^3_{j=1}(dx^j)^2 + b(t,y)^2 dy^2 \ .
\eea

Any time-dependence of the size of the fifth dimension is contained
in $b(t,y)$, since the boundary branes stay at constant coordinate position.

\paragraph{Brane matter.}

All brane matter is assumed to be created in a brane collision, so that in
the old scenario the hidden brane remains empty, while in the new scenario
the hidden brane may also contain matter. The brane matter action is
\bea \label{matteraction}
  S_{\textrm{matter}} = \sum_{i=1}^2 \int_{{\cal M}_4^{(i)}} d^4\xi_{(i)}\sqrt{-h_{(i)}}{\cal L}_{\textrm{matter}(i)} \ .
\eea

The detailed form of the matter Lagrange density is unimportant, since
under the assumptions of homogeneity and isotropy the energy-momentum
tensor of brane matter in any case has the ideal fluid form. Since the
treatment of brane matter in a cosmological context has thus far been
phenomenological, the brane matter term has not been included in the action
\re{hetaction}, even though five-dimensional heterotic M-theory does provide
a description of brane matter.

The calculational problem of the ekpyrotic scenario consists of obtaining
the brane interaction from M-theory, solving the equations of motion
obtained from \re{action} for the homogeneous and isotropic background using
initial conditions very near the BPS state, calculating the quantum
fluctuations of the branes as a perturbation of this background and finally
transferring the brane ripples into perturbations of brane energy
density. Even if the first step had been completed and the resulting terms
would be simple, solving the five-dimensional equations for the background
would be difficult. Therefore, the ekpyrotic scenario has mostly been
discussed in the framework of an effective four-dimensional theory.

\section{The four-dimensional effective theory} \label{4dtheory}

The four-dimensional effective theory of the ekpyrotic scenario is motivated
by the complexity of the five-dimensional equations, the ease at which one
can implement an effective treatment of the brane interaction in four
dimensions and the idea that at low energies there should exist an effective
covariant four-dimensional description of the locally observable physics. I
will first present the effective theory in this section and then discuss its
shortcomings in section \ref{problems}.

\subsection{The homogeneous and isotropic background}

The procedure of obtaining the four-dimensional effective theory from the
five-dimensional theory consists of two different approximations. The first
is the ``moduli space approximation''. The idea is that as long as the
system evolves slowly, the evolution can be described as movement in the
space of vacua spanned by the integration constants of the BPS solution. So,
one takes the BPS solution \re{BPS} and promotes the integration constants
$N, A, B, C$ and $Y$, known as ``moduli'', to functions which
depend on coordinates parallel to the branes, so that for the homogeneous
and isotropic background they depend only on time. In the moduli space
approximation, the metric and the fields are
\bea \label{moduli}
  ds^2 &=& -N(t)^2 D(t,y) dt^2 + A(t)^2 D(t,y)\sum^3_{j=1}(dx^j)^2 + B(t)^2 D(t,y)^4 dy^2  \el
  e^{\phi(t,y)} &=& B(t) D(t,y)^3 \el 
  {\cal F}_{0123y}(t,y) &=& -\alpha A(t)^3 N(t) B(t)^{-1} D(t,y)^{-2} \qquad\qquad y\le Y(t) \el
  & & - ( \alpha-\beta ) A(t)^3 N(t) B(t)^{-1} D(t,y)^{-2} \ \quad y\ge Y(t) \ ,
\eea

\noindent where $D(t,y)=\alpha y - \beta(y-Y(t))\theta(y-Y(t))+C(t)$;
$\theta(y-Y)$ is the step function.

The second approximation is to substitute the moduli metric and fields
\re{moduli} back into the action \re{action} and integrate
over the fifth dimension to obtain a four-dimensional action. The idea
behind this procedure is that as one cannot resolve the fifth dimension at
low energies due to its small size, one can integrate over it, a standard
prescription in the Kaluza-Klein approach to extra dimensions. Since the
dependence of the moduli metric and fields on the transverse coordinate $y$
factorises, the integration is trivial. To the resulting four-dimensional
action one then adds a potential term to support the movement in the space
spanned by the moduli. In the old ekpyrotic scenario the potential is for
the modulus $Y$, whereas in the new scenario it is presumably for $B$. As
mentioned earlier, these terms are hoped to be eventually computable from
heterotic M-theory.

The resulting four-dimensional action for the old ekpyrotic scenario is,
with the approximations $B=$ constant, $C=$ constant and with a small bulk
brane tension, $\beta\ll\vert\alpha\vert$ \cite{Khoury:2001a},
\bea \label{4daction}
  S_{\textrm{4d}} \approx 3 M_5^2 \int d^4 x \nt \at^3 \left( -\frac{1}{\nt^2}\frac{ \atdot^2}{\at^2} + \frac{\beta}{I_3}\left[ \frac{1}{2}\frac{1}{\nt^2} D(Y)^2 \Ydot^2 - \frac{V(Y)}{B I_3 M_5} \right] \right) \ ,
\eea

\noindent where $I_3(t)$ is a positive function which is constant to zeroth
order in $\beta/\alpha$, $V(Y)$ is the potential added by hand and
$\nt$ and $\at$ are defined in terms of the moduli $N$, $A$ and $B$ as
\bea \label{NandA}
  \nt &\equiv& N \sqrt{B I_3 M_5} \el
  \at &\equiv& A \sqrt{B I_3 M_5} \ .
\eea

The relation of the effective four-dimensional action of the new ekpyrotic
scenario to the five-dimensional action has not been presented, but is
presumably similar. The authors of the ekpyrotic scenario identify
\re{4daction} as the action of a four-dimensional homogeneous,
isotropic and spatially flat universe containing a scalar field
$Y$ that is minimally coupled to gravity. Apart from the appearance of
$D(Y)$ in the kinetic term of $Y$, the action has the standard form
in the approximation where one keeps only the leading terms in
$\beta/\alpha$ (in other words, neglects the time-dependence of $I_3$).

It would seem that a four-dimensional covariant low energy
effective theory has been obtained, though it should be immediately
emphasised that the lapse function $\nt$ and scale factor $\at$ are
not the lapse function and scale factor measured at the visible brane.
The scenario can then be analysed using the standard methods of general
relativity plus scalar fields, in the same manner as scalar field
inflation. Work on the ekpyrotic scenario has concentrated on such analysis,
especially on analysis of the perturbations around the homogeneous and
isotropic background. The analysis of the background proceeds as follows.

The effective potential for the brane distance is taken to be
\bea \label{potential}
  V(Y) = - F(Y) V_0 e^{- c Y} \ ,
\eea

\noindent where $V_0$ and $c$ are positive constants and $F(Y)$ takes into
account that the potential vanishes when the bulk brane collides with the
visible brane; it is assumed that $F(Y)=0$ for $Y=0$ and $F(Y)\approx1$
everywhere else. It is possible to use a potential that is not exponential,
as long the it satisfies $V V''/V'{}^2\approx1$. For example, a power law
potential $V(Y)\propto Y^q$ with a large $q\gtrsim40$ will also do
\cite{Khoury:2001a, Kallosh:2001a}.

The behaviour of a scalar field with the potential \re{potential} and
minimally coupled to gravity is simple to analyse in the homogeneous
and isotropic case. There is a solution where the scale factor
contracts obeying a power law, $\at\propto t^p$, with $p=2/(M_4^2 c^2)$.
The contraction of $\at$ was interpreted as the contraction of the
fifth dimension, which led to the problem of stabilising the collapse or
dealing with the boundary brane collision, which in turn led to the new
ekpyrotic scenario.

The new ekpyrotic scenario sprung from the idea that if the fifth dimension
is not stabilised and the boundary branes will eventually collide, this
collision can be used to ignite ekpyrosis, rendering the bulk brane
unnecessary. From the point of view of the four-dimensional effective
theory, the scenarios are quite similar. The main difference is that in the
old scenario, ekpyrosis occurs before the effective scale factor vanishes,
while in the new scenario the branes collide at that very moment. The branes
are then supposed to bounce apart, and the scale factor is supposed to start
expanding from zero. The interbrane distance is supposed to be eventually
stabilised via some as of yet unknown mechanism.

\subsection{Perturbations around the background}

As stressed in chapter \ref{cosmo}, a model of the primordial universe
should give quantitative predictions about the anisotropies of the CMB.
In order to be in agreement with observation, the temperature
fluctuations should be mostly adiabatic, nearly scale-invariant and have
an amplitude of about $10^{-5}$. In the ekpyrotic scenario the origin of
the CMB anisotropies is the quantum fluctuations of the interbrane
distance, which in the four-dimensional effective theory are treated
like the fluctuations of a scalar field. Perturbations 
have been extensively discussed in the four-dimensional effective theory
and in similar settings
[1, 2, 7, 113, 114, 120-134].

The perturbation theory around the collapsing background is well-known.
The perturbations are adiabatic since they are fluctuations of the same
quantity that is responsible for the collapsing background, and getting
the correct amplitude is simply a question of tuning some parameters.
The crucial question is whether the perturbations are scale-invariant.
At first sight this might seem not to be the case.

In inflation the perturbations are scale-invariant because both the Hubble
parameter and the scalar field are almost constant. In the original version
of the pre-big bang scenario with only one field, the perturbations have a
large spectral index, $n\approx4$, due to the rapidly decreasing Hubble
parameter. The effective four-dimensional theory of the ekpyrotic scenario
is much like the pre-big bang scenario, so one would expect a blue spectrum.
It seems that one does obtain a large spectral index, $n\approx3$, for some
perturbations. However, it is apparently also possible to get a spectral
index close to unity for some perturbation variables, by tuning the
potential \re{potential} to be very flat, $c\gg1$, so that the collapse is
very slow, $p\ll1$. The question is then: what is the perturbation variable
whose spectrum is inherited by the CMB fluctuations?

The only ambiguity in the evolution of the perturbations is due
to the curvature singularity where the scale factor vanishes. Since
the background is not well-defined at this point, it is clear that
perturbation theory around the background makes no sense either. However,
the approach taken by the authors of the ekpyrotic scenario is to find
perturbation theory variables which remain finite and match
these across the collision. Such a prescription requires that the singularity
is resolved in some manner which leaves the perturbation theory unaffected.
This problem is somewhat similar to the graceful exit problem of the
pre-big bang scenario, not least because it is also unresolved. The
resolution has been suggested to happen in the five-dimensional context
\cite{Khoury:2001c, Khoury:2001d, Seiberg:2002, Tolley:2002}, but no
consistent and detailed account has been given. The proposal in
\cite{Khoury:2001c, Khoury:2001d, Seiberg:2002, Tolley:2002} is
formulated in flat spacetime, and does not apply to the ekpyrotic
setting where brane tension will always curve spacetime \cite{Rasanen:2001}.

The matching across the ``bounce'' has been much debated
[7, 113, 114, 121, 122, 124-130, 133, 134].
It seems that, first, there is no unambiguous way to choose how to match
across the bounce, and second, that the impact on the CMB depends
sensitively on the matching conditions chosen. Furthermore, it has been
demonstrated that the treatment of the bounce as a sharp transition to match
across may not necessarily be justified \cite{Martin:2001}, not a surprising
result. It has also been argued that regardless of the matching, a
consistent large-scale treatment of the perturbations within the context of
general relativity will never yield the desired scale-invariant spectrum
\cite{Hwang:2002, Hwang:2001a}, and that perturbation theory breaks down
even before the singularity \cite{Lyth:2001a}. At best, the result of a
scale-invariant spectrum within the four-dimensional effective theory rests
on matching conditions which are, to quote one of the authors, ``a guess''
\cite{Turok:2001}.

The ekpyrotic scenario has been much criticised for the ambiguities
associated with the spectral index and for the prescription used in joining
the collapsing phase to an expanding one. However, a more fundamental issue
is that the whole framework of the four-dimensional effective theory is
highly questionable.

\section{Problems of the four-dimensional approach} \label{problems}

Some have pointed out \cite{Brandenberger:2001, Hwang:2001a, Notari:2002}
that the issues of bounce and perturbations should be properly handled in the
context of the five-(or higher)dimensional theory due to the singularity and
the associated ambiguities of the effective theory. However, the
four-dimensional description is problematic already at the homogeneous and
isotropic level, even without considering the singularity. I will now briefly
go through the problems of the four-dimensional approach. Some of the issues
were first brought up in \cite{Kallosh:2001b}, and others were highlighted in
\cite{Rasanen:2001}.

\subsection{The five-dimensional equations of motion}

The basis of the four-dimensional effective theory, the moduli space
ansatz \re{moduli} along with the potential for the interbrane
distance, does not satisfy the five-dimensional equations of motion
derived from the action \re{action} \cite{Kallosh:2001b, Rasanen:2001}.
As noted before, the potential was added directly to the four-dimensional
effective theory, and it was implied that in the five-dimensional theory it
would be a delta function source located on the moving brane. However, such
a delta-function potential cannot support time-dependent movement of the
bulk brane with the moduli metric: quite simply, the brane does not move.

Even if one adds an arbitrary energy-momentum tensor in the bulk, the
movement of the bulk brane is so limited as to render the ekpyrotic scenario
unworkable. The bulk brane can only move if the
brane interaction is coupled to the Calabi-Yau breathing modulus $\phi$,
and even then the velocity $\Ydot$ will vanish at the brane collision,
leading to zero ekpyrotic temperature, according to the original formulae
\cite{Khoury:2001a}.

The new ekpyrotic scenario does not necessarily suffer from such problems,
but since the curvature and energy density given by the moduli ansatz
diverge at the collision of boundary branes, it is clear that the moduli
space approximation does not work in the new ekpyrotic scenario, at least
near the collision\footnote{Note that it is possible to have non-singular
boundary brane collisions, given a less symmetric metric; see section
\ref{collision}.}.

These problems are simply a consequence of the overly constrained form of
the moduli metric \re{moduli}. In particular, the factorisation of the
$y$-dependence of the metric in the directions parallel to the brane is
known to be a severely constraining condition
\cite{Mohapatra:2000, Lesgourgues:2000}.

In this connection it may also be noted that the moduli metric cannot
support matter on the branes \cite{Enqvist:2001a}. The factorisable form of
the metric is so constraining that the Israel junction conditions which
relate the embedding of the brane to its energy density and pressure do not
permit any matter, as illustrated in section \ref{rscosmology}.

According to the authors of the ekpyrotic scenario, the moduli ansatz does
not need to satisfy the equations of motion. It is difficult to understand
how the moduli space approximation is supposed to work without satisfying
the dynamical equations of the theory even at some approximate level. Also,
it is unclear how one could evaluate the validity of the effective theory
except by comparing with the full theory, given that the
parameters that are supposed to remain small in order for the approximation
to be valid (time derivatives of physical quantities, presumably) can take
any value without degrading the internal consistency of the effective
theory. There is, for example, no expansion in terms of these parameters,
and it is not clear what the corrections to the moduli space approximation
would be.

But even if we took for granted that the moduli space approximation
does not need to satisfy the equations of motion and can describe
brane movement, and that the curvature singularity can be ignored,
the constraining form of the metric is responsible for other severe problems.

\subsection{Flow of energy off spacetime}

A reasonable condition for a theory formulated on a compact manifold
or orbifold is that no energy should flow away across the boundary of
spacetime. If the manifold or orbifold has no boundary, this
condition is trivially satisfied; otherwise it provides a non-trivial
boundary condition.

In the new ekpyrotic scenario, the ``no-flow condition'' for the
moduli metric \re{moduli} is violated whenever the hidden brane
moves \cite{Rasanen:2001}. Or, more reasonably, the no-flow condition
prevents the hidden brane from moving within the confines of the
moduli space ansatz.

The old ekpyrotic scenario fares little better. In the approximation $B=$
constant, $C=$ constant used in \cite{Khoury:2001a}, movement of the bulk
brane (as well as the boundary branes) is prohibited by the no-flow
condition. Relaxing these conditions on $B$ and $C$, it is not impossible
for the bulk brane to move without energy flowing away across the boundary
of spacetime. However, due to the constrained form of the moduli metric the
no-flow condition relates the bulk brane movement to the size of the fifth
dimension, making another problem apparent.

\subsection{The expansion of the fifth dimension}

The main problem of the old ekpyrotic scenario was considered to be
that the fifth dimension contracted as the bulk brane travelled from
the hidden brane to the visible brane. As noted above, the moduli
metric is so constrained that the no-flow condition relates the size
of the fifth dimension to the movement of the bulk brane. Setting
aside the problem that the bulk brane cannot actually move at all
with the potential used, or with $B$ and $C$ constant, let us see
what this relation shows. The velocity of the size of the fifth dimension
$L(t)\equiv\int_0^R dy\, b(t,y)=\int_0^R dy B(t) D(t,y)^2$ is \cite{Rasanen:2001}
\bea \label{Ldot}
  \Ldot(t) &=& \int_0^R dy\, \bdot(t,y) \el &=& -\frac{\beta\Ydot}{\beta Y + (\alpha-\beta) R} B R (\alpha Y + C)^2 \ .
\eea

Recall that $\vert\alpha\vert>\beta>0$, and that the bulk brane travels from
$y=R$ to $y=0$, so that $\Ydot<0$. Therefore, the sign of the velocity
$\Ldot$ is opposite to the sign of the tension of the visible brane,
$-\alpha$. In the set-up analysed in \cite{Khoury:2001a}, the tension of the
visible brane is negative, so that the fifth dimension expands, in
contradiction with the result of the four-dimensional effective theory that
it collapses. Let us recall that this collapse was regarded as the most
severe problem of the old ekpyrotic scenario.

\subsection{The Hubble law}

The contradictions between the results of the five-dimensional theory and
the four-dimensional effective theory are not confined to the bulk equations
of motion which are not satisfied or to the boundary conditions which lead 
to expansion of the fifth dimension. A comparison between the Hubble law 
observed on the visible brane obtained from the effective four-dimensional 
equations and the real Hubble law from the exact five-dimensional 
equations provides further illustration of the status of the effective 
theory.

The equations of the four-dimensional effective theory are
those of Einstein gravity plus a scalar field. However, as emphasised in
section \ref{4dtheory}, the parameters of the effective theory are not the
parameters seen by an observer on the visible brane (for example, the
contraction of the effective scale factor $\at$ does not imply contraction
of the scale factor of the visible brane). The Hubble law resulting from the
action \re{4daction} is given in \cite{Khoury:2001a} as
\bea \label{4dhubblea}
  \frac{1}{\at^2}\left(\frac{1}{\at}\frac{d\at}{d\eta}\right)^2 &=& \frac{\beta M_5}{B (I_3 M_5)^2}\left( \frac{1}{2} D(Y)^2  \left(\frac{dY}{d\eta}\right)^2 + V(Y) \right) \ ,
\eea

\noindent where $\eta$ is conformal time ($\nt=\at$). In the above 
equation it is assumed that $B=$ constant, $C=$ constant. Taking 
\re{4dhubblea} to be correct (though there should presumably be
$\at^{-2}$ multiplying the kinetic term) and changing to general time
($\nt$  undetermined), we obtain
\bea \label{4dhubbleb}
  \frac{1}{\nt^2}\frac{\atdot^2}{\at^2} &=& \frac{\beta M_5}{B (I_3 M_5)^2}\left( \frac{1}{2} \frac{\at^2}{\nt^2} D(Y)^2 \Ydot^2 + V(Y) \right) \ ,
\eea

Recall that $\nt$ and $\at$ are not the physical lapse function and scale
factor. Their relation to the components $n$ and $a$ of the physical metric
is, according to \re{moduli} and \re{NandA},
\bea \label{4dmetric}
  n(t,y) &=& \nt(t) \sqrt{ \frac{D(t,y)}{B I_3(t) M_5} } \el
  a(t,y) &=& \at(t) \sqrt{ \frac{D(t,y)}{B I_3(t) M_5} } \ ,
\eea

\noindent where $D(t,y)$ is given immediately after \re{moduli} and, as 
mentioned earlier, $I_3(t)\simeq I_3(0)+\mathcal{O}(\beta/\alpha)$. On the
visible brane we have
\bea \label{4dapprox}
  n(t,0) &\simeq& \nt(t) \sqrt{ \frac{C}{B I_3(0) M_5} } \el
  a(t,0) &\simeq& \at(t) \sqrt{ \frac{C}{B I_3(0) M_5} } \ ,
\eea

\noindent where terms of order $\beta/\alpha$ have been dropped.
Combining \re{4dhubbleb} and \re{4dapprox} we obtain the derivative
of the physical scale factor with respect to the physical proper time
measured on the visible brane ($n(t,0)=1$),
\bea \label{effhubble}
  \frac{\adot(t,0)^2}{a(t,0)^2} &\approx& \frac{\beta}{C I_3}\left( \frac{1}{2} a(t,0)^2 D(Y)^2 \Ydot^2 + V(Y) \right) \ ,
\eea

The Hubble law \re{effhubble} is to be compared with the Hubble law
that emerges from the exact five-dimensional equations derived directly from
\re{action} (assuming that the brane interaction has only a delta function
support at the bulk brane, and taking into account that no energy should
flow away across the boundary of spacetime) \cite{Enqvist:2001a}:
\bea \label{hethubble}
  \frac{\adot(t,0)^2}{a(t,0)^2} \,=\, -\frac{1}{6 M_5^3}\alpha e^{-\phi_1}\rho_r - \frac{1}{6 M_5^3}\alpha e^{-\phi_0}\rho_d + \frac{1}{36 M_5^6}\rho_m^2 + \frac{{\cal C}}{a(t,0)^4} \ ,
\eea

\noindent where $\phi_0$ is the constant value of the breathing modulus at
the position of the visible brane, and $\phi_1$ and ${\cal C}$ are some
constants. The tension of the visible brane has to be positive, $-\alpha>0$,
in order to recover a positive gravitational coupling constant, as noted in
section \ref{rsgravity}; no such restriction is apparent in the effective
four-dimensional theory. In the pre-collision era when the brane is empty,
the Hubble law \re{hethubble} reduces simply to
\bea
  \frac{\adot(t,0)^2}{a(t,0)^2} \,=\, \frac{{\cal C}}{a(t,0)^4} \ .
\eea

It is evident that the interbrane distance does not appear as a scalar
field, nor does the potential with delta function support enter the
equation at all. These are well-known general features of brane cosmology
\cite{Shiromizu:1999}, discussed in section \ref{branegravity}.

Let us also note that matter would presumably be added to the
four-dimensional effective theory in the same way as $V(Y)$.\footnote{As is
done in the cyclic model; see section \ref{cyclic}.} Then the gravitational
constant to which this matter couples would also not depend on the brane
tension but would be simply $M_5^{-2}$, according to the identification made
from \re{4dtheory}. There would also be no $\rho^2$-term in the effective
theory.

In summary, the Hubble law of the effective four-dimensional theory is
not only in quantitative but also in qualitative contradiction with
the real Hubble law given by the five-dimensional equations in at least
three important respects. In the real Hubble law 1) there is no scalar
field corresponding to the interbrane distance, 2) the gravitational
coupling has the same sign as the tension of the visible brane, and is not
given simply by $M_5^{-2}$ and 3) there is a term proportional to $\rho^2$.

\subsection{Integration over the fifth dimension} \label{integ}

As the above considerations amply demonstrate, the four-dimensional
effective theory might at best be called misleading. However,
one may ask whether the problems of the effective theory are
\emph{technical}, to be overcome with an improved method of
approximation --perhaps including a replacement of the moduli space ansatz
\re{moduli} with a less constrained configuration-- or whether they are
\emph{conceptual}, underlining a problem in the whole
approach of integrating along the fifth dimension to obtain a
four-dimensional theory.

Leaving the moduli space approximation aside, let us consider the other
ingredient: integration over the fifth dimension\footnote{The discussion in
this section draws heavily on \cite{Mennim:2000}.}. It is clear that such a
procedure is not covariant, since it singles out the direction transverse to
the branes. Without a metric which has a
simple factorisable form\footnote{As illustrated in section
\ref{rscosmology}, brane matter will always cause the metric to be
non-factorisable.}, the ``direction transverse to the branes'' is not
well-defined in the bulk: one may perform coordinate transformations that
mix the $t$- and $y$-coordinates in the bulk without affecting them on the
branes. So, it is not clear along which path to integrate. One might suggest
integrating along a geodesic transverse to the branes, but such a
prescription is clearly not unique: a geodesic which starts transverse to
the visible brane will not in general be transverse to the hidden brane, and
vice versa.

At any rate, as a matter of mathematical consistency one should not simply
sum (or integrate) tensor quantities (such as the metric) at different
points of the manifold but take proper care in transporting them to the same
point. It might seem that this is not a problem if one integrates at the 
level of the action --which is a scalar quantity-- as is done in the 
ekpyrotic scenario, instead of integrating over the equations of motion. 
However, in order to obtain the four-dimensional 
equations of motion from the four-dimensional action, one has to vary the 
action with respect to some four-dimensional quantity. In the ekpyrotic 
scenario this quantity is $\nt$, which is in no way covariant as is 
apparent from \re{4dmetric}, so that the mathematical status of the 
resulting equations is somewhat unclear.

Quite apart from such mathematical concerns, one may ask what is the
\emph{physical} justification for integrating over the fifth dimension. In
Kaluza-Klein theories, as well as in string theory, the idea is that the
observers are not localised in the extra dimensions, and so cannot resolve
them. Then it makes sense to integrate over the extra dimensions to obtain an
averaged theory. But that is not the situation here: the defining feature of
brane cosmology is that the observers \emph{are} localised in the extra
dimensions, being confined to a codimension one object -- in the simplest
case, one point along the extra dimension. Therefore, it does not seem to
make sense even from the physical point of view to integrate over the extra
dimensions to obtain an effective theory.

The authors of the ekpyrotic scenario have presented an additional
justification for using a four-dimensional effective theory. The idea has
also been used in the cyclic model, where there is no attempt to derive the
effective theory from a fundamental theory. The argument is that there
should be some covariant effective four-dimensional theory at low energy,
and that the simplest such theory is Einstein gravity plus a scalar field
with some potential. There are two counter-arguments to this proposition.

First, it is explicitly known in brane cosmology that one does not recover
Einstein gravity plus a scalar field at low energies. As discussed in section
\ref{branegravity}, if the brane does not have a tension that is positive,
one does not even approximately recover ordinary gravity. If the brane has
positive tension, one has Einstein gravity plus terms quadratic
in the brane energy-momentum tensor (the $\rho^2$-term) plus contributions
which cannot be solved from the brane equations, but have to be calculated
from the five-dimensional bulk equations. The description of the extra
dimension as a scalar field is a feature of Kaluza-Klein theories that does
not appear in brane cosmology, and likewise, the $\rho^2$-behaviour is nowhere
to be found in the effective four-dimensional theory or in Kaluza-Klein
theories  in general.

Second, in the four-dimensional effective theory (as formulated in
the ekpyrotic scenario and in the cyclic model), one does not
actually even recover Einstein gravity plus a scalar field with some
potential at low energy. In the ekpyrotic scenario, the physical
Hubble law given by \re{effhubble} is obviously not of that simple
form. Redefining the field $Y$ to obtain a canonical kinetic term
will result in a potential with a complicated dependence on an integral
over $D(Y) a(t(Y))$, certainly not the simplest form one
could imagine (though the inclusion of $\at^{-2}$ in the kinetic term
in \re{4dhubblea} would make the field redefinition a simple affair).
In the cyclic model, the departure from Einstein gravity plus scalar
field is even more apparent, as we will see in section \ref{cyclic}.

To summarise, the correct procedure for obtaining the physical Hubble rate
(and other parameters) in brane cosmology is not to integrate over the
fifth dimension but to consider the induced equations on the brane. These
equations do not in general form a closed system, meaning that one must
solve the full five-dimensional equations. This is not surprising. Were one
to consider a codimension one domain wall in the visible four-dimensional
universe, surely it would not be expected that the dynamics within the
wall could be solved entirely without reference to matter in the rest of the
universe.

\section{Problems in five dimensions}

In order to place the ekpyrotic scenario on a solid footing it is
necessary to study the full five-dimensional equations. The ekpyrotic
scenario is still young, and there has not been much work in that
direction, so instead of giving an account of results, I will only be
able to list some problems that a five-dimensional construction
should solve. Apart from \cite{Enqvist:2001a, Rasanen:2001}, the only
work on the issue is some criticism and discussion in
\cite{Kallosh:2001a, Kallosh:2001b} and some responses in
\cite{Donagi:2001, Khoury:2001b}.

Unlike in the four-dimensional effective theory, there are no
conceptual problems in the five-dimensional approach. The minimal action
of five-dimensional heterotic M-theory \re{hetaction} is well-established,
and the groundwork for the perturbation analysis in five-dimensional brane
cosmologies has also been done \cite{Perturbations}. The main
problem is that the brane interaction responsible for the breaking of the
BPS symmetry and the dynamics is not known in five dimensions, even at the
level of an effective description. However, there are some issues that can
be discussed without knowing the details of the brane interaction.

\subsection{Stabilisation of the fifth dimension} \label{stab}

Fitting tree-level parameters of (eleven-dimensional) heterotic M-theory to
the observed value of the gravitational coupling constant and the inferred
value of the grand unified gauge coupling, the size of the eleventh
dimension turns out to be a few orders of magnitude larger than the size of
the Calabi-Yau dimensions \cite{Horava:1995, Horava:1996}. It was this
observation that the universe would look five-dimensional over some energy
range that motivated the formulation of heterotic M-theory in five
dimensions \cite{Lukas:1998a, Lukas:1998b}. The size of the fifth dimension
is also important for the phenomenology of the model, as emphasised in the
ekpyrotic context in \cite{Kallosh:2001a, Kallosh:2001b}.

In the new ekpyrotic scenario the boundary branes collide together and
bounce apart, so that the problem of stabilising the brane distance at the
desired value is obvious. And even the old ekpyrotic scenario may have
problems with stabilisation, quite apart from the results of the
four-dimensional effective theory. The stabilisation mechanism of the
eleventh dimension is not known in heterotic M-theory. But since the brane
interaction that breaks the symmetry of the initial state is supposed to be
extremely weak, one may wonder whether a stabilisation mechanism would
interfere with the delicate journey of the bulk brane, or vice versa, a
point made in \cite{Kallosh:2001a, Kallosh:2001b}.

However, it is not clear whether such a stabilisation mechanism is needed.
The question is related to an important difference between a brane set-up
and a Kaluza-Klein set-up. In the original formulation of heterotic M-theory
the gravitational coupling depends on the size of the eleventh dimension
as in Kaluza-Klein models generally. However, as we have seen, that is
not true when one takes into account the brane nature of the dimensional
reduction from eleven to ten dimensions. I have not looked into the issue in
detail and do not know how the grand unified gauge coupling behaves in the
brane setting. The matter deserves further study, and the need for a
stabilisation mechanism in a brane setting should be carefully investigated.

\subsection{The boundary brane-boundary brane collision} \label{collision}

The new ekpyrotic scenario was originally presented because the effective
four-dimensional theory indicated that the fifth dimension collapses, so
that the eventual collision of the boundary branes rendered the bulk brane
superfluous. Even though there is no reason to take the four-dimensional
effective theory seriously, a formulation with only the boundary branes has
aesthetic appeal, as well as possibly being easier to solve. (There is also
the advantage of not having to address the question of the origin of the
bulk brane.) However, a collision between two boundary branes is a more
violent event than a collision between a boundary brane and a bulk brane,
since it involves one dimension vanishing for an instant. One might expect
the curvature to become singular at such a collapse, as happens with the
moduli metric, signalling the breakdown of the five-dimensional theory.

A proposal for resolving the issue has been made in \cite{Khoury:2001c,
Khoury:2001d, Seiberg:2002, Tolley:2002}. The matter has been investigated
in some detail in \cite{Rasanen:2001}. It turns out, perhaps surprisingly,
that it is possible for the transverse direction to collapse to a point and
re-expand without the curvature or the energy density becoming singular. This
is to be contrasted with a spatially flat or open Friedmann-Robertson-Walker
universe, where the simultaneous collapse of the three spatial dimensions
necessarily involves a curvature singularity\footnote{Some spatially closed
FRW universes can remain non-singular at the collapse, most notably the
3+1-dimensional Milne universe which contains no matter. However, the
reason for avoiding a singularity is different from the ekpyrotic case.}.
However, the non-singular configurations are very constrained. It is
possible to construct all non-singular metrics near the
collision and use the Israel junction conditions \re{israel} to look for
constraints on the matter created in the collision.

Since the brane interaction is supposed to vanish as the branes meet,
its energy-momentum tensor contributes only to terms in the Einstein
equation which remain bounded at the collision. Therefore the
energy-momentum tensor of brane interaction is not relevant for the
analysis of possible singularities at the collision, and the issue
may be studied on quite general grounds.

The basic requirement for a collision to be non-singular is that
physical quantities, in particular the Riemann tensor and the
energy-momentum tensor remain bounded, both in the bulk and on
the branes. In \cite{Rasanen:2001} the issue was studied in
a particular local orthonormal frame. (Un)boundedness of the
components of the Riemann tensor in a particular local orthonormal
frame does not guarantee their (un)boundedness in another
local orthonormal frame, since singularities may appear in local 
Lorentz transformations connecting different frames\footnote{I am
grateful to Jorma Louko for pointing this out.}. However, it is
straightforward to check that the constraints on brane matter given
below are valid in all local orthonormal frames which respect the
homogeneity and isotropy with respect to the spatial dimensions
parallel to the brane and reproduce the metric \re{metric}.

Other conditions necessary for consistency are the
no-flow condition and the covariant conservation of the
energy-momentum tensor.

The boundedness condition constrains the near-collision metric
significantly. The implications for brane matter depend
on how rapidly the fifth dimension re-expands. More specifically,
writing the scale factor of the fifth dimension in \re{metric}
after the collision as $b(t,y)=b_k(y) t^k+b_{k+1}(y) t^{k+1}+\mathcal{O}(t^{k+2})$, the implications for brane matter depend on the value of $k$
(which has to be an integer).

In the case $k=1$ in \cite{Rasanen:2001} there is both a sign mistake
and a term missing in equation (28) for $n$. The correct equation is
\bea
  \frac{1}{b_1^2} \left(n''_1 - \frac{b_1'}{b_1}n'_1\right) + n_1 &=& 2 \frac{b_2}{b_1} \ .
\eea

The inclusion of $b_2(y)$ is important, since it means that
the brane matter created in the collision is related not only to the
velocity $\bdot$ but also to the acceleration $\bddot$. The acceleration
at collision is not constrained by the boundedness requirement, so that
neither is the brane matter created.

In the case $k\ge2$ the acceleration does not enter and the brane matter
is severely constrained; it has to obey the relations
\bea
  \rho_1 + \rho_2 &=& 0 \el
  p_1 + p_2 &=& - 4 M_5^3 \delta_{2k} \int_0^R\! dy\, b_k(y)  \ ,
\eea

\noindent where $\rho_i$ and $p_i$ are the energy density and pressure,
respectively, of matter on brane $i$ immediately after the collision.
So, the energy density on one brane and the pressure on at
least one brane have to be negative.

It is worth emphasising that the above results follow from a direct
study of the Riemann tensor, and do not utilise the equations of
motion apart from the Israel junction conditions that give the embedding
of the branes into spacetime. (The no-flow condition of the Einstein tensor
does not yield new information.) Therefore, they are unaffected by changes
in the action unless they lead to a change in the junction conditions. Even
then, some constraints on brane matter (in the case $k\ge2$)
are likely to remain due to the highly constrained form of the
near-collision metrics.

In a way it is not surprising that one of the energy densities created on
the branes has to be negative (in the case $k\ge2$). According to
\re{hethubble} the
gravitational coupling on one brane will necessarily be negative. For the
standard four-dimensional Hubble law $H^2=8\pi G_N\rho_m/3$ this would imply
a non-positive energy density. Since the Hubble law is not the standard one,
this is not necessarily true, and the question would have to be
addressed in the context of a full solution of the five-dimensional
equations. However, it is clear that in the ekpyrotic brane setting one
cannot simply put ordinary matter on both branes without worrying about the
details, as originally assumed in the new ekpyrotic scenario.

For the minimal action of heterotic M-theory, \re{hetaction}, the no-flow
condition and the covariant conservation of the energy-momentum tensor
(taking into account an arbitrary brane interaction which vanishes at the
collision) forbid the possibility $k=1$, implying a negative energy
density. However, this result can possibly be avoided by turning on more
fields in the full action of five-dimensional heterotic M-theory.

To summarise, it is not impossible to have non-singular ekpyrotic boundary
brane collisions, but they are highly constrained. However, boundary brane
collisions which produce radiation on the visible brane are not ruled out.

\section{The ``cyclic model of the universe''} \label{cyclic}

Some of the ideas of the ekpyrotic scenario have been central in the
construction of a spin-off called the ``cyclic model of the universe''
[13-15]. The set-up is a
five-dimensional brane model, with matter produced in collisions between
boundary branes, as in the new ekpyrotic scenario. There are two main
differences between the cyclic model and the new ekpyrotic scenario.

The first important difference is that instead of being a unique
event, ekpyrosis is posited to occur at regular intervals. The
history of the universe then consists of an infinite sequence
of roughly identical cosmological cycles. Late-time inflation
serves to empty the branes between collisions, producing the highly
symmetric initial state postulated in the ekpyrotic scenario.

The second important difference is that even though the scenario is
motivated by heterotic M-theory, it is not based on it, or on any other
theory in the sense that the effective four-dimensional theory would be
derived from some more fundamental setting. Instead, the four-dimensional
theory --which is similar to that of the ekpyrotic scenario-- is simply
proposed ad hoc, on the principle of Einstein gravity plus scalar field
being the simplest possible covariant low energy description. The connection
to a more fundamental theory is hoped to eventually emerge. This connection
is especially important since the scale factor of the effective theory
collapses to a point, so that a higher-dimensional description is considered
necessary to resolve the apparent singularity. The issue has been discussed
in \cite{Tolley:2002}, and proposals in this direction made in the context
of the ekpyrotic scenario \cite{Khoury:2001c, Khoury:2001d, Seiberg:2002}
have also been referred to in the cyclic model.

Since the cyclic model is not based on a definite fundamental theory or a
given five-dimensional action, it is impossible to analyse it by starting
from the fundamental theory, as was done for the ekpyrotic scenario in
section \ref{4dtheory}. However, it is possible to make a few observations
based on general results in brane cosmology; the following remarks mostly
follow \cite{Rasanen:2001}.

\paragraph{The Hubble law.}

First, as in the ekpyrotic scenario, the physical Hubble law derived from
the effective theory is completely different from the real Hubble law in a
brane cosmology setting. The Hubble law on the negative tension brane in the
cyclic model, given by equations (8) and (12) of \cite{Steinhardt:2001b},
is\footnote{I have corrected a typo regarding $8\pi G$; it has no effect on
the present argument.}
\bea \label{cyclichubble}
  \frac{\adot_1^2}{a_1^2} &=& \frac{8\pi G}{3} \left(\beta^4 \rho + V(\phit)\right)  + 2 \phitdot \coth\phit \sqrt{\phitdot^2 + \frac{8\pi G}{3}\left(\beta^4 \rho + V(\phit)\right)} \el
  & & + \phitdot^2 ( 1 + \coth^2\phit) \ ,
\eea

\noindent where $a_1$ is the scale factor of the negative tension brane, $G$
is a constant, $\rho$ is the energy density of matter on the brane, $\phit$
is a scalar field related to the size of the fifth dimension\footnote{In
the notation of \cite{Steinhardt:2001b},
$\phit=(\phi-\phi_\infty)/\sqrt{6}$.}, $V(\phit)$ is the potential
responsible for brane movement and $\beta(\phit)=-2\sinh\phit$. Note that
the gravitational coupling is given by $8\pi G\beta^4$ and depends on the
size of the fifth dimension.

As an aside, the parameter with respect to which the time derivatives are
taken in \re{cyclichubble} is called the ``FRW proper time'' in
\cite{Steinhardt:2001b}. It is, however, apparently not the physical time;
for small brane separation the physical time is given in
\cite{Steinhardt:2001b} as $t_5=\int dt\, e^{-\phit}$. This detail makes
no difference to the present argument.

Since the higher-dimensional origin of the potential $V$ is not known, let
us consider the case $V=0$ to allow for comparison with the Einstein
equation which arises in the five-dimensional brane setting with an empty
bulk. The induced Einstein equation is given by \re{rshubble} with
$\Lambda=0$ and $K=0$,
\bea \label{realhubble}
  \frac{\adot^2}{a^2} \,=\, -\frac{\vert\Lambda_1\vert}{18 M_5^6}\rho + \frac{\Lambda_1^2}{36 M_5^6} + \frac{1}{36 M_5^6}\rho^2 + \frac{{\cal C}}{a^4} \ ,
\eea

\noindent where $\Lambda_1=-\vert\Lambda_1\vert$ is the brane tension.

A comparison of \re{cyclichubble} and \re{realhubble} shows that the
Hubble law of the cyclic model differs from the Hubble law given by
the five-dimensional equations in significant respects, as was the
case for the Hubble law of the ekpyrotic scenario. In the real Hubble law 1)
there is no scalar field corresponding to the interbrane distance, 2) the
gravitational coupling has the same sign as the tension of the visible
brane, and does not depend on the size of the fifth dimension, 3) there is a
term proportional to $\rho^2$ but no term involving $\rho$ under a square
root\footnote{As \re{inducedgen} shows, it is impossible to obtain such a
term in the five-dimensional brane setting without invoking an explicit
$\rho$-dependence in the bulk energy-momentum tensor.} and 4) there is a
term involving the square of the brane tension\footnote{In the
Randall-Sundrum model and in the ekpyrotic scenario this term is cancelled
by a bulk contribution; no such contribution has been specified in the
cyclic model.}.

In brief, the Hubble law of the cyclic model is completely different
from the Hubble law of the brane setting that it is supposed to describe.

\paragraph{The collision.}

Second, in \cite{Steinhardt:2001b} the spacetime is assumed to be
flat immediately before and after the collision, as argued in
\cite{Khoury:2001c, Khoury:2001d, Seiberg:2002}. However, brane
tension (and brane matter) necessarily implies that curvature
cannot be neglected \cite{Rasanen:2001}.

As noted in section \ref{collision}, non-singular boundary brane
collisions that produce radiation on the positive tension brane are
not ruled out. However, one cannot simply produce radiation on the
negative tension brane without additional complications, because of the
negative gravitational coupling.

\paragraph{The low energy effective description.}

Third, as is evident from \re{cyclichubble}, the aim of having the simplest
possible low energy description in terms of Einstein gravity plus a scalar
field is not realised, just like in the ekpyrotic scenario.

\paragraph{}

Problems that the cyclic model encounters even if one takes the
four-dimensional effective theory for granted have been discussed in
\cite{Linde:2002, Felder:2002}. It is also pointed out in
\cite{Linde:2002, Felder:2002} that, unlike the ekpyrotic scenario, the
cyclic model is not an alternative to inflation: the initial state which
solves the problems of homogeneity, isotropy and flatness is provided by
inflation instead of supersymmetry.

Apart from the above problems, the cyclic model is considerably less
attractive than the ekpyrotic scenario simply because it is not founded on a
fundamental theory. The effective theory is taken ad hoc, and the necessary
higher-dimensional description of the brane collision is adopted from a
study in flat spacetime \cite{Khoury:2001c, Khoury:2001d, Seiberg:2002, Tolley:2002} which is not applicable to the cyclic model because of the presence
of brane tension and matter \cite{Rasanen:2001}.

\section{Summary}

The ekpyrotic scenario is a promising concept. Starting, like the
pre-big bang scenario, from a set-up as fundamental (and therefore
speculative) as string/M-theory and descending down to
phenomenology is quite an attractive alternative to the
``bottom-up'' approach of most inflationary models. The possibility
of obtaining a non-singular cosmology is promising, and the
conceptual simplicity of the ekpyrotic scenario has definite appeal.

In brief, the ekpyrotic scenario is a welcome new idea.

However, the techniques employed so far in the study of the scenario are
mostly not solid, and more careful work is needed to promote the scenario
from a promising idea to a concrete model with testable predictions free
from technical problems. In particular, the four-dimensional effective
theory does not give a correct description, so the analyses within the
framework of this theory, including the work on perturbations
[1, 2, 7, 113, 114, 120-134]
are irrelevant as regards the ekpyrotic scenario.

Nevertheless, work on the four-dimensional effective theory has led
to renewed interest in collapsing cosmological backgrounds
[113, 125-128, 130, 132-134, 138, 139]
which may provide important insights, for example for the pre-big bang
scenario. Given the status of the effective four-dimensional theory,
it is ironic that the collapse problem seems to also have rekindled
serious interest in studying string theory in time-dependent
backgrounds, which may be a first step towards a resolution of real
cosmological singularities
[135, 140-142].

Ultimately, the form of the brane interaction, the absorption of the bulk
brane by the visible brane and the small instanton phase transition will
have to be addressed in the M-theory context if the ekpyrotic scenario is to
be a fundamental description of the early universe. However, it might be
possible to do some useful analysis within the context of the
five-dimensional effective theory. A promising route might be to consider
the bulk brane case, construct a reasonable ansatz for the brane interaction
and solve the equations of motion for the background. Then one should do the
perturbation analysis, building on existing techniques in brane cosmology
\cite{Perturbations} and construct an effective description of transferring
the brane ripples to perturbations of brane energy density.

The ekpyrotic scenario is only one year old, and the next few years
will show whether it leads to a working model that solves cosmological
problems or whether its main impact will be as a springboard for new ideas.

\end{document}